\documentclass{emulateapj}

\def\ltsima{$\; \buildrel < \over \sim \;$}
\def\simlt{\lower.5ex\hbox{\ltsima}}
\def\gtsima{$\; \buildrel > \over \sim \;$}
\def\simgt{\lower.5ex\hbox{\gtsima}}

\shorttitle{The Chemical Composition of Sgr}
\shortauthors{McWilliam, Wallerstein, \& Mottini}

\begin{document}

\topmargin 0.5in

\newcommand{\znh}{[{\rm Zn/H}]}
\newcommand{\msol}{M_\odot}
\newcommand{\etal}{et al.\ }
\newcommand{\delv}{\Delta v}
\newcommand{\kms}{km~s$^{-1}$ }
\newcommand{\cm}[1]{\, {\rm cm^{#1}}}
\newcommand{\N}[1]{{N({\rm #1})}}
\newcommand{\e}[1]{{\epsilon({\rm #1})}}
\newcommand{\f}[1]{{f_{\rm #1}}}
\newcommand{\rAA}{{\AA \enskip}}
\newcommand{\sci}[1]{{\rm \; \times \; 10^{#1}}}
\newcommand{\ltk}{\left [ \,}
\newcommand{\ltp}{\left ( \,}
\newcommand{\ltb}{\left \{ \,}
\newcommand{\rtk}{\, \right  ] }
\newcommand{\rtp}{\, \right  ) }
\newcommand{\rtb}{\, \right \} }
\newcommand{\ohf}{{1 \over 2}}
\newcommand{\nohf}{{-1 \over 2}}
\newcommand{\rhf}{{3 \over 2}}
\newcommand{\smm}{\sum\limits}
\newcommand{\perd}{\;\;\; .}
\newcommand{\cmma}{\;\;\; ,}
\newcommand{\intl}{\int\limits}
\newcommand{\mkms}{{\rm \; km\;s^{-1}}}
\newcommand{\ew}{W_\lambda}


%
%
%
%
%
\title{Chemistry of the Sagittarius Dwarf Galaxy: a Top-Light IMF, Outflows and the R-Process}

\author{Andrew McWilliam$^1$}
\affil{The Observatories of the Carnegie Institute of Washington, \\
813 Santa Barbara Street, Pasadena, CA 91101}
\email{$^1$ andy@obs.carnegiescience.edu} 

\and 

\author{George Wallerstein$^2$ and Marta Mottini}
\affil{Astronomy Department, University of Washington, Seattle, WA 98195}
\email{$^2$ walleg@u.washington.edu}

\begin{abstract}

From chemical abundance analysis of stars in the Sagittarius dwarf
spheroidal galaxy (Sgr), we conclude that the alpha-element deficiencies 
cannot be due to the Type~Ia supernova (SNIa) time-delay
scenario of Tinsley (1979).  Instead, the evidence points to low
[$\alpha$/Fe] ratios resulting from an initial mass function (IMF) deficient in
the highest mass stars.  The critical evidence is the 0.4
dex deficiency of [O/Fe], [Mg/Fe] and other hydrostatic elements, contrasting
with the normal trend of r-process [Eu/Fe]$_r$ with [Fe/H].  
Supporting evidence comes from the hydrostatic element (O, Mg, Na, Al, Cu)
[X/Fe] ratios, which are inconsistent with iron added to the Milky Way (MW) disk
trends.  Also, the ratio of hydrostatic to explosive (Si, Ca, Ti)
element abundances suggests a relatively top-light IMF.
Abundance similarities with the LMC, Fornax and IC~1613, suggest that
their alpha-element deficiencies also resulted from IMFs lacking the most
massive SNII.

For such a top-light IMF, the normal trend of r-process [Eu/Fe]$_r$ with [Fe/H],
as seen in Sgr, indicates that massive Type~II supernovae ($\simgt$30M$_{\odot}$)
cannot be major sources of r-process elements.

High [La/Y] ratios, consistent with leaky-box chemical evolution, are confirmed
but $\sim$0.3 dex larger than theoretical AGB predictions.  This may
be due to the $^{13}$C pocket mass, or a difference
between MW and Sgr AGB stars.  Sgr has the lowest
[Rb/Zr] ratios known, consistent with low-mass ($\simlt$2M$_{\odot}$) AGB stars near
[Fe/H]=$-$0.6, likely resulting from leaky-box chemical evolution.

The [Cu/O] trend in Sgr and the MW suggest that
Cu yields increase with both metallicity and stellar mass, as
expected from Cu production by the weak s-process in massive
stars.

Finally, we present an updated hfs line list, an abundance
analysis of Arcturus, and further develop our error analysis formalism.

\end{abstract}

\section{Introduction}

The Sagittarius dwarf Spheroidal galaxy (Sgr), which is merging with the Milky Way (MW), has
six associated globular clusters: Ter~7, Pal~12, Whiting~1, M54, Arp~2 and Ter~8.  The
first three clusters are middle-aged (6--8 Gyr) and metal-rich, with [Fe/H] between
$-$0.6 and $-$0.8 dex, while the last three are old (11--14 Gyr)  and metal-poor, 
with [Fe/H] in the range
$-$1.6 to $-$2.3 dex.  M54 (=NGC~6715), the most populous cluster, lies in the densest part
of Sgr (Ibata et al. 1994; Sarajedini \& Layden 1995; Monaco et al. 2005).  This has
prompted discussion of whether M54 is the nucleus of the Sgr galaxy (Sarajedini \& Layden 1995;
Da Costa \& Armandroff 1995; Layden \& Sarajedini 2000), around which later star formation occurred.

Ground-based photometric observations by Sarajedini \& Layden (1995), Layden \& Sarajedini (2000), 
and Hubble Space Telescope photometry of M54 by Siegel et al. (2007) shows that M54 contains a second,
fainter, red giant branch (RGB).  
Analysis by Siegel et al. (2007) indicated an old, 13 Gyr, population with the presence of at least
two intermediate-aged populations: 4--6 Gyr with [Fe/H]=$-$0.4 to $-$0.6 dex, from the fainter
RGB, and a 2.3 Gyr population near solar abundance.
These observations may indicate that M54 has two, or more, populations with significantly
different [Fe/H], making it an unusual globular cluster, similar to Omega Cen.   
The density of stars in the faint RGB Sgr population is larger in the 
central M54 field than the Sgr field, studied by Layden \& Sarajedini (2000), which
suggests the possibility that the faint RGB stars belong to a late episode of star formation
within the globular cluster, rather than the Sgr galaxy itself.  
Monaco et al. (2005) found a strong density enhancement in Sgr, with a peak that is
indistinguishable from the center of M54; they concluded that Sgr is a nucleated dwarf galaxy,
even when M54 is ignored.


However, Siegel et al. (2007) claim that the fainter RGB belongs to Sgr, and that it is
possible that M54 formed separately from Sgr's nucleus and was later pulled into the galaxy's 
center through dynamical friction.

A spectroscopic study by Bellazzini et al. (2008) found that the velocity dispersion of
the faint RGB is identical to that of Sgr, which is constant from 0.5 arc minutes out to 
at least 
100 arc minutes from M54.  This constant velocity dispersion for Sgr, at 9.6 km/s, over an extended
region, much larger than M54, is strong evidence supporting the idea that the faint RGB in M54
is part of the Sgr galaxy, rather than the globular cluster.  This differs from the velocity
dispersion trend in M54, which declines from 14.2 km/s at the center to $\sim$5.0 km/s at a distance
of 3.5 arc minutes.

Bellazzini et al. (2008) also found a bimodal metallicity function, based on Ca-triplet
measurements, one peaked at [Fe/H]=$-$1.45 and the other at [Fe/H]=$-$0.45, in good agreement
but $\sim$0.1 dex higher than other metallicity estimates for the main body of M54 and 
Sgr, respectively.  Thus, Bellazzini et al. (2008) showed that the faint RGB towards
M54 is, in fact, due to Sgr.

Several photometric and spectroscopic studies of M54 have been undertaken, providing information
on the cluster's characteristics (distance, reddening, etc.) and estimates of the metallicity.
Sarajedini \& Layden (1995) found [Fe/H]=$-$1.79$\pm$0.08 from photometry of RGB stars, and an
intrinsic dispersion of 0.16 dex.  Da~Costa \& Armandroff (1995) estimated M54's metallicity 
from the Ca~II triplet, assuming that the [Ca/Fe] ratio is similar to normal metal-poor stars,
and found [Fe/H]=$-$1.55$\pm$0.10 dex.  Sarajedini \& Layden (1997) discussed the systematic
effects on metallicity estimates from photometry and Ca~II triplet spectroscopy in the case
on non-standard [$\alpha$/Fe] ratios.

The first high-resolution model atmosphere abundance analysis of M54 stars was undertaken by
Brown, Wallerstein \& Gonzalez (1999).  They studied five red giants  belonging to the bright, 
or main, RGB of M54, and found [Fe/H]=$-$1.55$\pm$0.10 dex, with oxygen and other
alpha-elements characteristic of the Milky Way (henceforth MW) halo.

%

From 2MASS photometry of stars in the Sgr red giant branch (RGB), Cole (2001) found a
mean [Fe/H] of $-$0.5$\pm$0.2 dex.  Detailed, high resolution, abundance studies of Sgr
stars include Bonifacio et al. (2000, 2004; henceforth B00, B04), 
Smecker-Hane \& McWilliam (2002; henceforth SM02), Sbordone et al. (2007; henceforth S07)
and Carretta et al. (2010; henceforth C10).  These studies found 
[Fe/H] for individual Sgr stars, ranging from below $-$1 to above solar. 
In their study, C10 found [Fe/H]=$-$1.56 dex for M54, based on 76 stars, and
a mean [Fe/H]=$-$0.62 for Sgr from 27 stars.  

In this paper we discuss a detailed chemical abundance analysis of three red giant stars 
belonging to the fainter RGB toward M54, which  Bellazzini et al. (2008) found to be
kinematic members of Sgr.  We compare our results to previous abundance studies of Sgr,
in order to understand the complex chemical evolution of this system.

In \textsection 2 we describe our observations and the data reduction, in \textsection 3 we
describe the abundance analysis procedures, and we discuss our findings and conclusions in 
\textsection 4, 5 and 6.


\section{Observations and Data Reduction}

We acquired high resolution spectra of three stars on the fainter giant branch
of M54 in July 2007.  Target selection employed the color-magnitude diagrams and
finding chart for the M54 shallow frames of Layden \& Sarajedini (2000).  The stars 
were chosen to be relatively bright 
(V$\sim$16.2), isolated, with colors indicating that these were late K-type giants, not M giants.  
This was important, because the TiO absorption present in M giants would have made the abundance
analysis significantly more difficult.  

The spectra were obtained using the Magellan Echelle spectrograph, MIKE, with a 0.5 arc sec
slit, corresponding to R$\sim$48,000.  Typical exposure times were 6000 sec, resulting in typical
S/N values near 40 at the peak of the H$_{\alpha}$ order; actual
exposure times and final, per extracted pixel, S/N values are listed in Table~\ref{tab-obs}.
Due to line crowding and reduced S/N toward the blue, the portion of our spectra
useful for abundance analysis ranged from 5120 to 9250\AA .

The spectra were reduced using the MIKE pipeline, written by Dan~Kelson (see Kelson 2003
for details).  In order to flatten the spectra we traced the continuum of a hot star,
HR9098 (B9IV), from a high S/N spectrum obtained in the same observing run
as the M54/Sgr stars.  The continuum fit for this {\em blaze standard} was performed
using the IRAF {\em continuum} routine, with a high order cubic spline fit.

\begin{deluxetable}{ccccccc}[h]
\tabletypesize{\scriptsize}
\tablecaption{Observations}
\tablewidth{0pt}
\tablehead{\\
\colhead{Star} & 
\colhead{R.A. (2000)} &
\colhead{Dec. (2000)} &
\colhead{V}&  
\colhead{Exp. (s)} &  
\colhead{S/N} & 
\colhead{RV\rlap{$_{\rm helio}$}} \\
}
\startdata
242   & 18:55:17.9 & $-$30:27:49 & 16.225 & 5700 & 39 & 145.4 \\
247   & 18:54:48.3 & $-$30:26:39 & 16.241 & 8100 & 50 & 145.4 \\
266   & 18:54:40.5 & $-$30:26:49 & 16.292 & 6000 & 42 & 131.8 \\
\enddata
\tablecomments{Star identifications and V magnitudes are from the shallow M54 frame data of
Layden \& Sarajedini (2000).  The S/N values are final, per extracted pixel, at the peak
of the H$_{\alpha}$ order}
\label{tab-obs}
\end{deluxetable}

\section{Abundance Analysis}

Radial velocities of our three stars were determined by measuring the central wavelengths
of a handful of strong lines in the spectra.  Conversion to heliocentric velocities was
accomplished using the IRAF {\em rvcorrect} algorithm; the results are listed in 
Table~\ref{tab-obs}.

The mean heliocentric velocities of our three stars is $+$140.9$\pm$4.6 km/s.
This value is consistent with measurements of the mean heliocentric velocities of both the 
M54 core and the Sagittarius nucleus, at $+$140.7$\pm$0.4 and $+$139.9 $\pm$0.6 km/s, respectively 
(see Bellazzini et al. 2008).
The velocity dispersion of our stars, at $\sigma$=7.9 km/s, is also consistent with the
constant value of $\sigma$=9.6 km/s for Bellazzini's Sagittarius nucleus, and their positions
in the color-magnitude diagram. 

In this work we employ equivalent width (EW) model atmosphere abundance analysis for
most elements; however, spectrum synthesis profile matching calculations were 
used to determine abundances in a few cases where critical lines were blended or
partially blended.  The lines used here
for abundance analysis were selected from a number of our previous papers on abundances
in red giant stars, including
Fulbright, McWilliam \& Rich (2007), Koch \& McWilliam (2008), 
Smecker-Hane \& McWilliam (2002, henceforth SM02) and McWilliam et al. (1995).  We also 
include a number of new lines identified as potentially useful in this work, in particular
Rb~I 7948\AA\ and the Zr~I lines at 8070, 8133, and 8389 \AA .

Because of the relatively low S/N of our spectra, especially at bluer
wavelengths, the lines we have used tend to be near the flux peaks of the echelle orders,
where relatively weak lines can be reliably measured.

Our line EWs were measured from the flattened spectra by AM and GW, independently,
using the IRAF {\em splot} routine.  To identify continuum regions and for blend detection
the high quality spectral atlas of Arcturus (Hinkle et al. 2000) was particularly useful.
The final, average, EWs for the lines used in this paper are listed in Table~\ref{tab-linesew}.

\begin{deluxetable*}{rccrcrcrcl}
\tabletypesize{\scriptsize}
\tablecaption{Line List}
\tablewidth{0pt}
\tablehead{ 
\colhead{Ion} & 
\colhead{$\lambda$} &  
\colhead{E.P.} &  
\multicolumn{2}{c}{Star 242} &
\multicolumn{2}{c}{Star 247} &
\multicolumn{2}{c}{Star 266} &
\colhead{notes} \\
\colhead{} & 
\colhead{[\AA ]} &  
\colhead{[eV]} &  
\multicolumn{1}{c}{EW} &
\colhead{$\Delta\varepsilon_{\rm aboo}$} &
\multicolumn{1}{c}{EW} &
\colhead{$\Delta\varepsilon_{\rm aboo}$} &
\multicolumn{1}{c}{EW} &
\colhead{$\Delta\varepsilon_{\rm aboo}$} &
\colhead{} }
\startdata
[O I]& 6300.30 & 0.00  &  60  & $-$0.46   &        61   & $-$0.23       &       63  & $-$0.40 & ss,ew \\
\\
Na I & 6154.23 & 2.10  &  68  & $-$0.41   &        80   & $-$0.22       &       75  & $-$0.30 \\
Na I & 6160.75 & 2.10  &  80  & $-$0.60   &        86   & $-$0.37       &       98  & $-$0.22 \\
\\
Mg I & 5711.09 & 4.35  & 136  & $-$0.51   &       137   & $-$0.20       &      140  & $-$0.37 \\
Mg I & 6318.77 & 5.11  &  46\rlap{:} & $-$0.45   &        52\rlap{:}  & $-$0.23\rlap{:}      &      ...  &  ... \\
Mg I & 6319.24 & 5.11  &  34  & $-$0.42   &        40   & $-$0.21       &       28\rlap{:} & $-$0.53\rlap{:} \\
Mg I & 8717.82 & 5.93  &  56  & $-$0.45   &        64   & $-$0.14       &       52  & $-$0.49 \\
Mg I & 8736.02 & 5.96  &  94  & $-$0.44   &       119   & $+$0.17       &      100  & $-$0.32 \\
\enddata
\tablecomments{This table is published in its entirety in the 
electronic edition of the {\it Astrophysical Journal.} A portion
is shown here for guidance regarding its form and content.}
\label{tab-linesew}
\end{deluxetable*}

The abundance analysis follows the differential method, relative to Arcturus, we devised in
Fulbright et al. (2007) and employed by Koch \& McWilliam (2008, 2010, 2011).  We use the
spectrum synthesis program MOOG (Sneden 1973) and Kurucz' model atmospheres from his web site,
at {\tt http://kurucz.harvard.edu}, as discussed in Castelli, Gratton \& Kurucz (1997).

Line-by-line differential abundance studies possess several advantages that improve the accuracy of
abundance measurement, particularly when the target and standard star are of similar
spectral type. For example, the accuracy of the gf values becomes unimportant, thus
allowing lines with poorly measured or unknown gfs to be used.  Even the effects of unidentified 
line blends should be reduced when taking differential abundances, relative to a similar standard star.

Unaccounted effects in the
model atmospheres, such as 3D hydrodynamics, granulation, non-LTE, and the effect of a
chromosphere on the T-$\tau$ relation, are likely to be very similar in standard and target
stars, provided that the atmosphere parameters are close enough.  However, no calculations yet
exist to show how similar the program and standard stars need to be for good cancelation
of systematic errors.
On the other hand, so long as the model atmosphere abundance corrections from the various
unaccounted physical effects have the same sign in the program and standard stars, then the 
diffential abundances should be more reliable than the absolute model atmosphere abundances.


While use of the sun as a differential standard would always eliminate the gf value
problem, Arcturus should be a superior differential standard for our Sgr RGB target stars.
In particular, we can use lines that are present in Arcturus and our Sgr RGB
stars that are not detected in the solar spectrum (e.g., Rb~I 7948\AA\ and Zr~I
lines at 8070, 8133, and 8389\AA ).  

\begin{deluxetable}{lrrr}
\tabletypesize{\scriptsize}
\tablecaption{Arcturus Abundance Ratios}
\tablewidth{0pt}
\tablehead{\\
\colhead{Ion} &
\colhead{{[}X/Fe{]}}  &
\colhead{$\sigma$} &
\colhead{N\rlap{$_{\rm lines}$}} }\\
\startdata
{[}Fe\,I/H{]}               & $-$0.49 &  0.07 & 152    \\   
{[}Fe\,II/H{]}              & $-$0.40 &  0.04 &   8    \\

{[}O\,I{]}\rlap{a} & $+$0.46 &  ...  &   1    \\
Na\,I           	    & $+$0.09 &  ...  &   1    \\  
Mg\,I           	    & $+$0.39 &  0.06 &   5    \\  
Al\,I           	    & $+$0.38 &  0.03 &   3    \\  
Si\,I           	    & $+$0.35 &  0.05 &  15    \\  
Ca\,I           	    & $+$0.21 &  0.01 &   2    \\  
Ti\,I           	    & $+$0.26 &  0.04 &  17    \\  
Ti\,II\tablenotemark{a}     & $+$0.26 &  0.04 &   7    \\  
V\,I\tablenotemark{b}       & $+$0.12 &  0.02 &   3    \\
Mn\,I\tablenotemark{b}      & $-$0.14 &  0.09 &   4    \\
Cu\,I\tablenotemark{b}      & $+$0.30 &  ...  &   1    \\
Rb\,I\tablenotemark{b}      & $+$0.05 &  0.01 &   2    \\
Zr\,I\tablenotemark{b}      & $-$0.28 &  0.09 &   3    \\
Y\,II\tablenotemark{ab}     & $-$0.22 &  0.14 &   6    \\
Ba\,II\tablenotemark{a}     & $-$0.17 &  ...  &   1    \\
La\,II\tablenotemark{ab}    & $-$0.08 &  0.09 &   6    \\

Eu\,II\tablenotemark{ab}    & $+$0.23 &  0.06 &   2    \\
\\

[$\alpha$/Fe]               & $+$0.32 &  0.09 &   6\tablenotemark{c}    \\

%
\enddata
\tablenotetext{a}{Relative to \ion{Fe}{2}.}
\tablenotetext{b}{hfs treatment employed to compute abundances}
\tablenotetext{c}{Number of elements used to compute [$\alpha$/Fe]}
\tablecomments{Note that sigmas only reflect agreement within the line list for each species and do not
                 include systematic errors}
\label{tab-abooab}
\end{deluxetable}

However, to use Arcturus as a standard requires that we have an accurate measurement
of its chemical abundance distribution, relative to the sun.  Fortunately, an extremely
high quality spectrum of Arcturus, from UV to IR, is available (i.e., Hinkle et al. 2000),
matching the quality of the Kurucz et al. (1984) solar spectrum.  Thanks to the proximity of
Arcturus, its atmosphere parameters are known to a precision better than any other RGB 
star (see Fulbright et al. 2007; Koch \& McWilliam 2008).  

%
%

%
%
%
%

In this work we increase the number of elements in Arcturus with differential
abundance ratios, [X/Fe], relative to the sun, using the same
Arcturus model atmosphere as Koch \& McWilliam (2008).  The adopted abundance ratios
are presented in Table~\ref{tab-abooab}.

\subsection{Hyperfine and Isotopic Splitting}

For most species we employed the single line, EW, line-by-line, differential 
abundance analysis, relative to Arcturus.  However, the lines of a 
number of elements studied here suffer from de-saturation due to hyperfine splitting and/or
isotopic splitting (hereafter hfs).  For these elements it was necessary to include the fine structure
components in the spectrum synthesis calculations.  Where possible, we employed hfs lists
that we had previously employed or calculated in other studies, but for a few
lines we searched the literature for the latest hfs and isotopic splitting constants.
Thus, our hfs line lists are the best available for several lines, in particular the
Rb~I and Zr~I lines.  The hfs energy level
splittings were computed using equations 1 and 2, below.  For Cu, Rb, Zr, and Eu we adopted
solar isotopic compositions in the line lists.  Table~\ref{tab-hfs} in the electronic version shows the
complete hfs and isotopic line lists used in this work.  In Table~\ref{tab-hfsrefs} of Appendix~II
we provide the references for the constants used to generate the hfs line lists in this work.

\begin{equation}
\Delta E = \frac{1}{2} AK - B \frac{\frac{3}{4}K(K+1)-J(J+1)I(I+1)}{2I(2I-1)J(2J-1)}
\end{equation}
where
\begin{equation}
K = F(F+1)-I(I+1)-J(J+1)
\end{equation}

Here, I is the nuclear angular momentum quantum number for the isotope, J is the total
electronic quantum number for the level, F is the total angular momentum quantum number
for the atom in a given hyperfine level (F is the vector sum of the nuclear and
electronic momenta) and A and B are the hyperfine constants.  The relative strengths
of the levels are computed according to Condon \& Shortly (1935, page 238).
Also, see Woodgate (1980) for a useful discussion.  We also recommend Johnson et al. (2006)
as a resource for hfs line lists.

\begin{deluxetable}{lcc}[!h]
\centering
\tabletypesize{\scriptsize}
\tablecaption{HFS List}
\tablewidth{0pt}
\tablehead{\\
\colhead{Species} & \colhead{Wavelength (\AA )} & \colhead{log F$_i$gf} } \\
%
\startdata
$^{87}$Rb\,I & 7800.2480  &  $-$1.2181 \\
$^{87}$Rb\,I & 7800.2515  &  $-$1.2181 \\
$^{87}$Rb\,I & 7800.2529  &  $-$1.6160 \\
$^{85}$Rb\,I & 7800.2993  &  $-$0.8855 \\
$^{85}$Rb\,I & 7800.3008  &  $-$0.7886 \\
$^{85}$Rb\,I & 7800.3013  &  $-$0.9013 \\
$^{85}$Rb\,I & 7800.3584  &  $-$0.4241 \\
$^{85}$Rb\,I & 7800.3608  &  $-$0.7886 \\
$^{85}$Rb\,I & 7800.3623  &  $-$1.3326 \\
$^{87}$Rb\,I & 7800.3813  &  $-$0.7709 \\
$^{87}$Rb\,I & 7800.3867  &  $-$1.2181 \\
$^{87}$Rb\,I & 7800.3901  &  $-$1.9171 \\
\\
$^{87}$Rb\,I & 7947.5747  &  $-$1.2249 \\
$^{87}$Rb\,I & 7947.5918  &  $-$1.9239 \\
$^{85}$Rb\,I & 7947.6323  &  $-$0.7954 \\
$^{85}$Rb\,I & 7947.6401  &  $-$1.3395 \\
$^{85}$Rb\,I & 7947.6963  &  $-$0.8923 \\
$^{85}$Rb\,I & 7947.7041  &  $-$0.7954 \\
$^{87}$Rb\,I & 7947.7188  &  $-$1.2249 \\
$^{87}$Rb\,I & 7947.7358  &  $-$1.2249 \\
\enddata
\tablecomments{F$_i$ indicates isotopic fraction.
This table is published in its entirety in the 
electronic edition of the {\it Astrophysical Journal.} A portion
is shown here for guidance regarding its form and content.}
\label{tab-hfs}
\end{deluxetable}

\subsection{Model Atmosphere Parameters}

Our stellar atmosphere parameters were determined using the VI photometry of
Layden \& Sarajedini (2000) and the JHK 2MASS data from Skrutskie et al. (2006).
Reddening corrections were based on the extinction relations of Winkler et al. (1997),
with E(B$-$V)=0.15, adopted from Siegel et al. (2007).  Color temperatures were based
on the calibration of Ram\'irez \& Mel\'endez (2005; henceforth RM05), assuming [Fe/H]=$-$0.5.  
This temperature scale is similar to other currently popular calibrations: the Casagrande
et al. (2010) scale is hotter than RM05 by 40K; RM05 is hotter than Alonso et al. (1999)
by 18K.  The physical temperature for Arcturus, from Fulbright et al (2006) is 4290K, with
a 1$\sigma$ uncertainty of $\pm$29K (Koch \& McWilliam 2008; henceforth KM08).  

The Lee (1970) and Johnson et al. (1966) photometry of Arcturus gives (V$-$I)$_J$, 
(V$-$J)$_J$, and (V$-$K)$_J$ of 1.62, 2.08 and 2.925 mag. respectively.  Note
that the 2MASS JHK photometry for Arcturus is highly uncertain, with 1$\sigma$$\sim$0.17 mag,
likely due to saturation effects, and therefore not used.  Transformation of the Johnson
(V$-$I) color to the Kron-Cousins system is accomplished by use of the relations in
Bessell (1979).  Transformation of the Johnson (V$-$J) and (V$-$K) colors to the
K-short system used by 2MASS was obtained by first converting from Johnson to CIT system
with the relations of Elias et al. (1985) and then to the 2MASS system with the relations
given by 
Carpenter (2005, unpublished).\footnote{http://www.astro.caltech.edu/\~jmc/2mass/v3/transformations/.}


We find that the RM05 V$-$K calibration gives T$_{\rm eff}$ for Arcturus of 4275K, some 15K cooler
than the physical effective temperature; this is well within the uncertainties on the physical
T$_{\rm eff}$ and corresponds to a mere 0.023 mag error in the V$-$K color.  Therefore, we
have adopted the RM05 photometric temperature calibration for this work.  
%
%
%
Table~\ref{tab-photteff} summarizes the photometry and resultant temperatures for each color,
based on the RM05 calibrations and [Fe/H]=$-$0.5; we also include the transformed colors
and color-temperatures derived for Arcturus, showing good agreement with the physical T$_{\rm eff}$.

\begin{deluxetable*}{ccccccccc}
\centering
\tabletypesize{\scriptsize}
\tablecaption{Photometry and Temperatures}
\tablewidth{0pt}
\tablehead{ 
\colhead{Star} &
\colhead{V} & 
\colhead{\quad(V$-$I)\rlap{$_0$}\quad\quad}& 
\colhead{(V$-$J)\rlap{$_0$}}& 
\colhead{(V$-$K)\rlap{$_0$}}& 
\colhead{T(V$-$I)}&
\colhead{T(V$-$J)}& 
\colhead{T(V$-$K)}& 
\colhead{$\overline{\rm T}$ \rlap{$_{\rm eff}$} }
}
\startdata
 242 &  16.225 & 1.604 & 2.707 & 3.663 & 3921 & 3916 & 3922  & 3920  \\  
 247 &  16.241 & 1.676 & 3.050 & 3.728 & 3873 & 3787 & 3897  & 3852  \\
 266 &  16.292 & 1.642 & 2.722 & 3.615 & 3895 & 3909 & 3941  & 3915  \\
Arcturus & $-$0.05 & 1.260 & 2.137 & 2.940 & 4272 & 4283 & 4275 & \\
\enddata
\tablecomments{De-reddened optical, V and Kron-Cousins I-band
photometry, from Layden \& Sarajedini (2000), and infrared photometry
from the 2MASS catalog (Skrutskie et al. 2006).  Reddening corrections
were based on Winkler et al.  (1997), with E(B$-$V)=0.15, adopted from
Siegel et al. (2007).  Arcturus colors transformed from Lee (1970) and
Johnson et al. (1966), see text.
Color temperatures were based on the calibration
of Ram\'irez \& Mel\'endez (2005), assuming [Fe/H]=$-$0.5.}
\label{tab-photteff}
\end{deluxetable*}

Photometric gravities were found using the adopted T$_{\rm eff}$ values and a 5 Gyr
(recommended by Siegel et al. 2007) Teramo canonical, scaled-solar composition, isochrone
with z=0.008.  We assumed that our stars are on the RGB; log\,g for AGB stars would have
been 0.07 dex smaller.

As usual, we iterated on the microturbulent velocity, metallicity 
and [$\alpha$/Fe] enhancement.  The microturbulent velocities were chosen by requiring 
that Fe~I abundances be independent of EW.  

Once the [Fe/H] derived from iron lines and model metallicity were
roughly consistent we computed the mean [$\alpha$/Fe] from our abundances of O, Mg, Si,
Ca and Ti, which we employed to select the appropriate [$\alpha$/Fe] ratio
of the model atmospheres for the abundance analysis.  The use of model atmospheres
with the appropriate [$\alpha$/Fe] ratio is necessary due to the
contribution of free electrons from the ionization of Mg and Si, which in turn affects the H$^-$ 
continuous opacity and computed line strengths, particularly for lines from species of the
dominant ionization stage (e.g., O~I, Fe~II, La~II, etc.).
Since our results show [$\alpha$/Fe] near zero, and below, for all three program stars, we
have adopted the scaled solar composition Kurucz model atmospheres.  Future analyses
might consider use of sub-solar [$\alpha$/Fe] model atmospheres for Sgr stars.
Our final adopted model atmosphere parameters are listed in Table~\ref{tab-params}.  

\begin{deluxetable}{ccccccc}
\tabletypesize{\scriptsize}
\tablecaption{Adopted Model Atmosphere Parameters}
\tablewidth{0pt}
\tablehead{ 
\colhead{Star} & 
\colhead{T$_{\rm eff}$} & 
\colhead{log\,$g$}  & 
\colhead{[A/H]}  & 
\colhead{$\xi$}  &
\colhead{[Fe\,I/H]\rlap{$_{\rm lines}$} } 
}
\startdata
 242 &  3920 & 0.96 & $-$0.5 & 1.7 & $-$0.49 \\
 247 &  3850 & 0.83 & $-$0.2 & 1.4 & $-$0.09 \\
 266 &  3920 & 0.93 & $-$0.5 & 1.6 & $-$0.39 \\
 Arcturus &  4290 & 1.60 & $-$0.5 & 1.6  & $-$0.49 \\
\enddata
\tablecomments{Gravities were found using the adopted T$_{\rm eff}$ values and a 5 Gyr
(recommended by Siegel et al. 2007) Teramo canonical, scaled-solar composition, isochrone
with z=0.008.  For log\,g determination stars were assumed to be on the RGB; log\,g for
AGB stars would have been 0.07 dex smaller.  Note that the observed M$_{\rm v}$ for Star
242 is closer to the AGB value than the RGB value in the Teramo isochrone.}
\label{tab-params}.
\end{deluxetable}

\begin{figure}[ht]
\centering
\includegraphics[width=8.0cm]{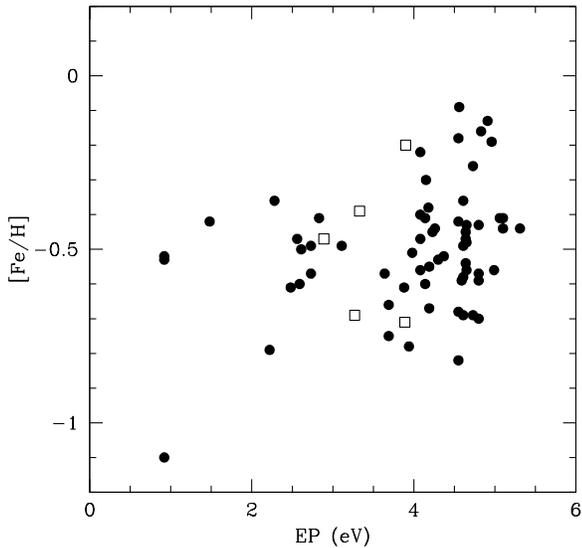}
\caption{Star \#242 [Fe/H] versus Excitation Potential, in eV, for differential line by line
Fe I (filled circles) and Fe~II (open squares) abundances.  The lack of a strong slope
indicates consistency of the line excitation with the adopted model atmosphere T$_{\rm eff}$,
determined photometrically.  Note that ionization equilibrium is obtained.
}
\label{fig-tex}
\end{figure}

A check on our photometric temperatures is obtained 
from a plot of differential Fe abundance versus line excitation potential, as shown in
Figure~\ref{fig-tex}.  We note that our Arcturus [Fe~I/H] and [Fe~II/H] values, taken
line by line relative to the sun are $-$0.49 and $-$0.40 respectively, a result similar
to that found by Koch \& McWilliam (2008).  The ionization imbalance might be due to
improper accounting for the electron number density, such as might occur with erroneous
[$\alpha$/Fe], or incorrect adopted gravity, or non-LTE over-ionization of Fe~I, or other
difficulties.  
For our stellar [Fe~I/H] and [Fe~II/H] abundances we are ultimately referenced to the
sun, so the Arcturus zero-point does not affect our results.  
For ionized species in the
program stars, we take the [X~II/Fe~II] ratios, which cancels-out the Arcturus
zero-point offset in Fe~II.

In principle, it should be possible to check the adopted gravity of our program stars
from ionization equilibrium of Fe and Ti, since Fe~II and Ti~II lines are sensitive to
the electron density in the atmosphere, which is strongly affected by gravity.  Unfortunately,
the large dispersion of our measured Ti~II abundances excludes this element for use as a
gravity discriminator.  For iron we note that while star~242 shows excellent agreement between
Fe~I and Fe~II abundances, the Fe~II abundances are 0.11 and 0.13 dex higher than
Fe~I in stars 247 and 266 respectively.  The excess Fe~II abundances could, reasonably, be due
to measurement error.  It cannot result from these two
stars being on the AGB rather than the RGB, since the small gravity change, of 0.07 dex,
leads to an abundance difference of only 0.03 dex (c.f. Table~\ref{tab-dabunddpar}); thus, 
a much larger gravity difference is required to explain the apparent difference between Fe~I 
and Fe~II abundances.  An alternative explanation, perhaps more realistic, for the ionization
imbalance is that the electron density, N$_e$, in stars 247 and 266 is significantly lower than
expected from the scaled solar model atmospheres employed due to the large underabundances
of Na, Mg, Al, and Si (see Table~\ref{tab-abunds}); these elements are important electron donors in
the atmospheres of our stars.  A calculation of N$_e$ for two locations in the line-forming 
region of the model atmosphere for star~266 showed that the measured abundance deficiencies
of O, Na, Mg, Al, Si, Ca and Ti lead to a reduction of N$_e$ by $\sim$30\%, or $\sim$0.1 dex,
roughly consistent with the putative ionization imbalance.
Other possible explanations for the apparent ionization imbalance include excessive mass-loss 
in the program stars, leading to lower than expected gravity; and strongly enhanced He abundances.

\section{Abundance Results and Discussion}

Abundance results for our three Sgr stars are provided in Table~\ref{tab-abunds}.  In this
table we list the [Fe/H] derived from Fe~I and Fe~II lines separately; we also show
element to iron ratios, [X/Fe], where the normalizing iron abundance (Fe~I or Fe~II)
is chosen to minimize the effects of systematic errors in the atmosphere parameters.
For lithium only, Table~\ref{tab-abunds} provides the absolute lithium abundances on the
hydrogen = 12.0 scale, $\epsilon$(Li); the abundances are based on the hfs line list of
Andersen et al. (1984) and are {\bf not} differential to any standard star.

\begin{deluxetable*}{lrrrrrrrrrrr}
\tabletypesize{\scriptsize}
\tablecaption{Abundance Results}
\tablewidth{0pt}
\tablehead {\colhead{} & \multicolumn{3}{c}{\#242} & & \multicolumn{3}{c}{\#247} & & \multicolumn{3}{c}{\#266} \\
\cline{2-4}\cline{6-8}\cline{10-12}
 \raisebox{1.5ex}[-1.5ex]{Ion} & \colhead{{[}X/Fe{]}} & \colhead{$\sigma$}& \colhead{N} & & \colhead{{[}X/Fe{]}} &
 \colhead{$\sigma$}& \colhead{N} & & \colhead{{[}X/Fe{]}} & \colhead{$\sigma$}& \colhead{N}}
\startdata
{[Fe\,I/H]}                 & $-$0.49 &  0.18 &  64 & & $-$0.09 &  0.19 &  65 & & $-$0.39 &    0.22 & 56 \\
{[Fe\,II/H]}                & $-$0.47 &  0.21 &   5 & & $+$0.02 &  0.12 &   5 & & $-$0.26 &    0.18 &  3 \\

{[O\,I]}\tablenotemark{a} & $-$0.02 &  ...  &   1 & & $-$0.19 &  ...  &   1 & & $-$0.04 &    ...  &  1 \\

Na\,I                       & $-$0.43 &  0.13 &   2 & & $-$0.63 &  0.11 &   2 & & $-$0.28 &    0.06 &  2 \\
Mg\,I                       & $-$0.07 &  0.03 &   5 & & $-$0.14 &  0.17 &   5 & & $-$0.15 &    0.10 &  4 \\
Al\,I                       & $-$0.09 &  0.07 &   5 & & $-$0.31 &  0.09 &   5 & & $-$0.09 &    0.12 &  2 \\
Si\,I                       & $-$0.08 &  0.16 &   7 & & $+$0.00 &  0.16 &   9 & & $+$0.08 &    0.22 &  9 \\
Ca\,I                       & $+$0.01 &  0.16 &   6 & & $-$0.17 &  0.16 &   4 & & $+$0.09 &    0.15 &  5 \\
Ti\,I                       & $-$0.08 &  0.18 &   9 & & $-$0.09 &  0.18 &   9 & & $-$0.02 &    0.14 &  9 \\
Ti\,II\tablenotemark{a}     & $+$0.00 &  0.20 &   2 & & $-$0.06 &  0.16 &   2 & & $-$0.14 &    0.37 &  2 \\

V\,I\tablenotemark{b}       & $-$0.08 &  0.08 &   3 & & $-$0.06 &  0.02 &   3 & & $+$0.09 &    0.10 &  3 \\
Mn\,I\tablenotemark{b}      & $-$0.27 &  0.08 &   4 & & $-$0.06 &  0.10 &   4 & & $-$0.03 &    0.05 &  4 \\
Cu\,I\tablenotemark{b}      & $-$0.64 &   ... &   1 & & $-$0.31 &   ... &   1 & & $-$0.44 &     ... &  1 \\
Rb\,I\tablenotemark{b}      & $-$0.19 &  0.01 &   2 & & $-$0.44 &  0.06 &   2 & & $-$0.21 &    0.02 &  2 \\
Zr\,I\tablenotemark{b}      & $-$0.08 &  0.10 &   6 & & $+$0.28 &  0.07 &   6 & & $+$0.08 &    0.08 &  6 \\
Y\,II\tablenotemark{ab}     & $-$0.20 &  0.10 &   5 & & $-$0.39 &  0.20 &   5 & & $-$0.08 &    0.13 &  5 \\
La\,II\tablenotemark{ab}    & $+$0.49 &  0.16 &   4 & & $+$0.48 &  0.23 &   4 & & $+$0.40 &    0.10 &  4 \\
Eu\,II\tablenotemark{ab}    & $+$0.39 &  ...  &   1 & & $+$0.23 &  ...  &   1 & & $+$0.32 &    ...  &  1 \\
\\

[$\alpha$/Fe]               & $-$0.04 &  0.04 &   6 & & $-$0.11 &  0.07 &   6 & & $-$0.03 &    0.10 &  6 \\
\\
log\,$\epsilon$(Li)\tablenotemark{c} & $+$0.42 & ... & 1 & & $+$0.14 &  ...  &   1 & & $-$0.47 &    ...  &  1 \\
\enddata
\tablenotetext{a}{Relative to \ion{Fe}{2}.}
\tablenotetext{b}{hfs/ew treatment employed to compute abundances}
\tablenotetext{c}{Absolute abundances listed for Li, based on the line list of Andersen et al. (1984) }
\tablecomments{Note that sigmas indicate the rms dispersion of the measurements for species with more
                 than one line; they are dispersions, not errors on the mean.}
\label{tab-abunds}
\end{deluxetable*}

\subsection{Iron}

The [Fe/H] values for our three stars, 242, 247 and 266, are $-$0.49, $-$0.09
and $-$0.39 dex respectively.  The dispersion of these [Fe/H] values is too
large to be due to measurement error about a single mean value for Sgr.  Our formal
best estimate for the internal 1$\sigma$ [Fe/H] measurement uncertainty is 0.03 dex
(see Appendix, Table~\ref{tab-finalerrors}), while our most pessimistic estimate of
this internal [Fe/H] uncertainty, within our sample,
is 0.10 dex.  The reduced Chi-squared fit assuming a single [Fe/H] value is
$\chi$$^2$$\sim$48 for the former measurement error, and $\chi$$^2$$\sim$4 for
the pessimistic 0.10 dex measurement uncertainty.  Thus, we conclude that our measurements
are best represented with a 1$\sigma$ intrinsic abundance spread, near 0.20 dex,
about the mean of [Fe/H]=$-$0.32 dex.  The conclusion that our reported [Fe/H] differences 
are real is supported by the systematic difference in line EWs between the three stars, 
even though the stellar temperatures are similar.

The [Fe/H] values of our three stars are consistent with the mean values and ranges of
photometric and spectroscopic metallicities reported for Sgr from a number of 
studies: Cole (2001), SM02, S07,
Siegel et al. (2007), Bellazzini et al. (2008), B00, B04 and C10.  
These studies show a consistent picture, with Sgr stars ranging in [Fe/H] from
near $-$1 to above the solar value, with a mean at approximately $-$0.5 dex.

While the mean [Fe/H], and [Fe/H] dispersion, found here are consistent with the results
of previous Sgr studies, our selection of stars from the faint RGB, below the M54 giant branch,
has biased us to the mean and more metal-rich side of the Sgr metallicity distribution.


\subsection{The Alpha Elements}

Wallerstein (1962) discovered that metal-poor, MW halo, stars showed excesses of
Mg, Si, Ca and Ti, relative to Fe; later, Conti et al. (1967) found similar
excesses for O.  These even-numbered elements (O, Mg, Si, Ca and Ti) have come to
be known as alpha elements, even though no single nuclear reaction produces their
excesses.

Historically, the decline in [O/Fe] (and other alpha elements) versus [Fe/H] from the MW
halo to disk has been assumed to result from the time delay between Type~II and Type~Ia supernovae
(hereafter SNII and SNIa), first detailed by Tinsley (1979). In this scenario, oxygen is produced
first by the short-lived SNII, while iron is produced by both SNII and SNIa; the
SNIa contribution occurred on longer timescales than the SNII.  In this way, SNIa iron was added
after a time delay for SNIa onset.

This SNIa time-delay chemical evolution scenario was explored and supported by
the detailed calculations by Matteucci \& Brocato (1990), which
predicted that the decline to lower [O/Fe] in systems with low star
formation rates (henceforth SFR), like the LMC,
should occur at lower [Fe/H] than the MW disk.  Similarly, the [O/Fe] decline should occur
at higher [Fe/H] in high SFR systems, like giant elliptical galaxies and bulges.


The average [$\alpha$/Fe], derived from our measured O, Mg, Si, Ca, and Ti abundances, is 
$-$0.04 dex, $-$0.11 dex, and $-$0.03 dex for stars 242, 247 and 266, respectively 
(see Table~\ref{tab-abunds}).  The average [$\alpha$/Fe], at $-$0.06 dex, is close
to the solar value, consistent with the scaled-solar model atmospheres
employed for the abundance analysis.  However, it is notable that the
most [Fe/H]-rich star, 247, has the lowest [$\alpha$/Fe] ratio, at $-$0.11 dex.

\begin{figure}[hb]
\centering
\includegraphics[width=8.0cm]{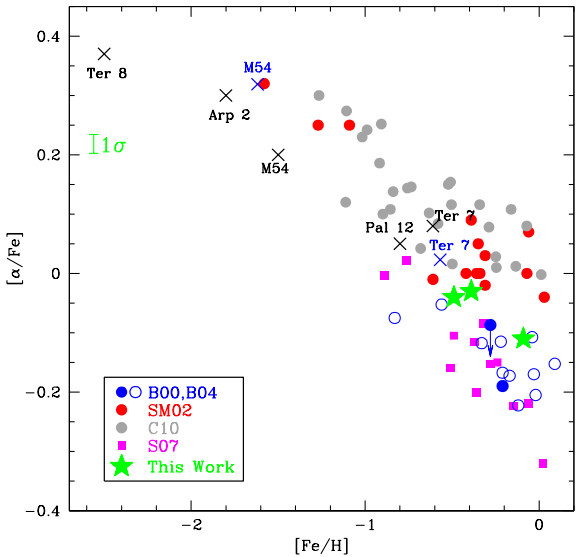}
\caption{[$\alpha$/Fe] ratios in Sgr.  Crosses indicate globular clusters
associated with Sgr: Terzan~8 and Arp~2 (Mottini et al. 2008), M54 (Brown et al. 1999; 
Carretta et al. 2010), Pal~12 (Cohen 2004), and Ter~7 (Tautvai\v{s}ien\.{e} et al. 2004;
Sbordone et al. 2007).  Sgr field stars are indicated for previous works:
filled gray circles (Carretta et al. 2010; C10), filled red circles (Smecker Hane \& McWilliam 2002; SM02), 
filled and open blue circles (Bonifacio et al. 2000, 2004; B00, B04) and filled magenta squares
(Sbordone et al. 2007; S07).
The Sgr field stars studied in this work are represented by filled green stars.  }
\label{fig-alphas}
\end{figure}

Figure~\ref{fig-alphas} shows a comparison of the average Sgr [$\alpha$/Fe] ratios found
here with other works.
The [$\alpha$/Fe] ratios measured by SM02
and C10 agree remarkably well. 
Results for stars 242 and 266 here are consistent with these papers, although on the lower envelope;
however, [$\alpha$/Fe] for star 247 (the most Fe-rich star) is lower by $\sim$0.1 dex.
We note that the spatial location of the stars studied here and those in SM02 and C10 possess 
considerable overlap, with roughly the same mean position within Sgr.

The S07 [O/Fe] and [Si/Fe] trends are similar to that found
by C10.  On the other hand, S07 found sub-solar [$\alpha$/Fe] values for their sample
of 12 Sgr stars, lower than other studies by 0.1--0.3 dex, with the
difference increasing to higher [Fe/H].  This deficiency in S07 is dominated by unusually
low [Ti/Fe] and quite low [Ca/Fe] and [Mg/Fe] abundance ratios;  these
sub-solar [Ca/Fe]
ratios are not seen in the MW disk at any metallicity.  Only star 247 from our sample
has sub-solar [Ca/Fe].  
Therefore, either there is a systematic error in the S07 [$\alpha$/Fe] values or a different 
composition for the S07 Sgr field, which is located 22 arc minutes West of the
stars in this work, SM02 and C10.  Clearly, further investigation of the low
alpha-element abundances found by S07 is warranted.

We note that a long-known disparity between Ti~I and Ti~II abundances may have contributed to
the low [Ti/Fe] ratios in S07, and other papers.  The cause of the Ti ionization imbalance has
recently been shown to be due to non-LTE effects on Ti~I, by Bergemann (2011).   Bergemann (2011)
finds Ti~I non-LTE corrections of $+$0.25 dex are required for the very metal-poor RGB
star HD~122563 ([Fe/H]=$-$2.5); whether the correction would be similar for solar-metallicity
RGB stars is not known.  Although substantial absolute non-LTE abundance corrections
for Ti~I are indicated, the line-by-line differential LTE abundance analysis of bulge
and disk giants by Fulbright et al. (2007) found
Ti-ionization imbalance of only $\sim$0.05 dex.  Presumably, the non-LTE effect in
the Fulbright program stars were cancelled-out by use of Arcturus as a differential
standard.  Our results show excellent agreement between Ti\,I and Ti\,II measurements,
with a mean $\varepsilon$(Ti\,I)$-$$\varepsilon$(Ti\,II) abundance difference
of 0.00 dex and an rms scatter of 0.10 dex.
Thus, by employing a differential line-by-line analysis we believe that we have
minimized systematic problems with non-LTE corrections to the Ti~I abundances.



Our [$\alpha$/Fe] ratios are slightly higher than, but on the upper envelope of, the results
of B00 and B04; notably, B04 did not measure Ti abundances for their stars.  Thus, our
[$\alpha$/Fe] points lie close to the average of C10/SM02 and S07/B00/B04.
The [$\alpha$/Fe] values for the Sgr fields near M54, studied here and by SM02 and C10,
are lower than the solar neighborhood thin-disk trend by $\sim$0.18 dex.  


The [O/Fe] ratio in our three stars is $\sim$0.3 dex below the trend in the solar neighborhood
measured by Allende Prieto et al. (2004); however, our [O/Fe] ratios are only 0.18 dex
below the solar neighborhood disk trend found by Edvardsson et al. (1993).  The difference
between these two comparison is that in the latter the sun is at the upper end of the
distribution of [O/Fe] ratios at [Fe/H]=0.0, while in the former study the sun has an 
unusually low [O/Fe] for its metallicity.  Thus, the difference lies in the systematic
effects present
in Edvardsson et al. (1993) and Allende Prieto et al. (2004).  Unsurprisingly, our measured
[O/Fe] ratios are consistent with McWilliam \& Smecker-Hane (2005a, henceforth MS05), which are deficient
relative to the solar neighborhood by 0.17 dex.  The [O/Fe] deficiencies
in C10 are slightly lower, by as much as 0.10 dex, than the results found in this work.  
The S07 [O/Fe] results are also $\sim$0.1 dex lower than found here,
similar to their generally low [$\alpha$/Fe] ratios.
In the SNIa time-delay paradigm of Tinsley and Matteucci \& Brocato the low [O/Fe] ratios indicate a lower
SFR in Sgr than the solar neighborhood.

%




\subsection{Lithium}


Our Li abundances, based on the 6707\AA\ line, are $+$0.42, $+$0.14 and $-$0.47 dex
for stars 242, 247 and 266 respectively. 
Although the EWs of the Li~I line in stars 242 and 247 are quite large, at 137 and 103 m\AA ,
this is due to the cool stellar temperatures rather than high Li abundances.

Because of their similar age ranges and metallicities it is sensible to compare the Li abundances
of our Sgr stars with Li abundances for red giant stars in the Galactic disk.
A comparison of our Li abundances with the LTE abundance survey
of solar neighborhood GK giants by Brown et al. (1989) shows that our stars 
fall near the peak of the Li-abundance frequency distribution function, and thus appear
quite normal.  We calculate that our three stars fit the distribution of Li abundances
in Brown et al. (1989) with $\chi^2$=1.10.  We note that the mean and standard deviation
of the Li-detections in Brown et al. (1989) are 0.48 and 0.69 dex, respectively, while
the mean and standard deviation of our three stars is 0.03 and 0.46 dex, respectively.
Clearly, it will be necessary to measure Li abundances in a larger sample of Sgr stars
before systematic differences between Sgr and the MW disk can be detected.
For now, it appears that whatever causes a range of Li abundance in solar neighborhood 
giants also applies to the Sgr RGB stars.  

Extensive non-LTE calculations for Li have been performed (e.g., Carlsson et al. 1994)
that suggest Li abundance corrections to the LTE values near $+$0.25 dex for our stars,
so the maximum non-LTE Li abundance for our stars is $+$0.6 dex.
However, we note that the Brown et al. (1989) sample has considerable overlap with our stars
in [Fe/H] and T$_{\rm eff}$.  Thus, any non-LTE corrections to the Li abundances 
for our Sgr stars are likely to be similar to those for local RGB stars in 
Brown et al. (1989).    Again, our stars show the normal range of Li abundances found for
GK giants in the Galactic thin disk.

\subsection{Sodium and Aluminum}

\begin{figure}[h]
\centering
\includegraphics[width=8.0cm]{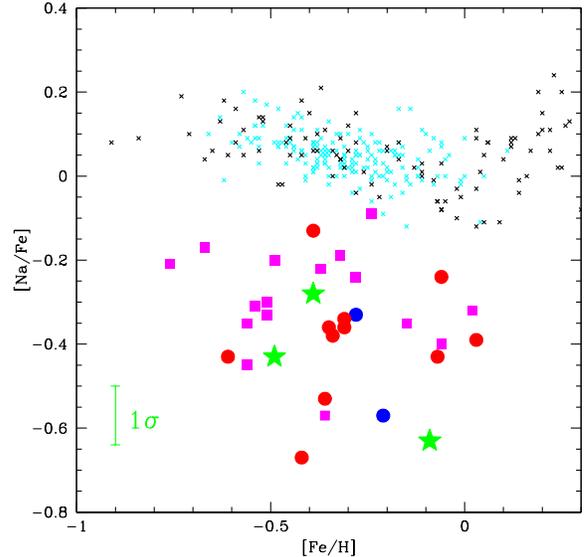}
\caption{[Na/Fe] for our program stars (filled green stars) compared to the Galactic thin disk
measurements of Bensby et al. (2005; black crosses) and Reddy et al. (2003; cyan crosses).
Sgr results from SM02 are represented by filled red circles,
filled blue circles indicate B00, and filled magenta squares represent S07.
Our program stars show the $\sim$0.4 dex [Na/Fe] deficiencies seen in Sgr by SM02,
S07, and Bonifacio et al. (2000); such deficiencies are seen in other dwarf galaxies by various authors.}
\label{fig-nafe1}
\end{figure}

In Figure~\ref{fig-nafe1} we compare the [Na/Fe] ratios found here
with the Galactic disk studies of Reddy et al. (2003) and Bensby et al. (2005), and
with the Sgr work of SM02, S07 and B00.  We find excellent agreement
with these previous Sgr works.  In all four investigations the Sgr stars are deficient
relative to the solar neighborhood stars, and all, except S07, are consistent with
$\sim$0.4 dex deficiencies (S07 found an average deficiency of $\sim$0.3 dex).  
We note that 0.3--0.4 dex deficiencies in [Na/Fe] are commonly seen in abundance studies of
other dwarf spheroidal galaxies (e.g., Shetrone et al. 2001, 2003).  

\begin{figure}[h]
\centering
\includegraphics[width=8.0cm]{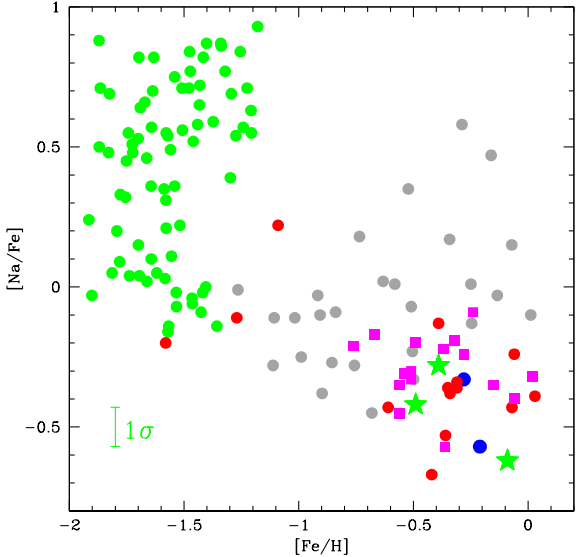}
\caption{
Larger scale plot of [Na/Fe] versus [Fe/H] for Sgr/M54. Filled green stars indicate our program stars.
Also shown are results from C10 for M54 (green filled circles) and Sgr dSph 
(gray filled circles).  Red filled circles are from SM02; magenta squares show S07 points;
blue circles are from B00.
Note that [Na/Fe] in Sgr found by S07 and B00 agree with this work, while
C10 [Na/Fe] ratios are higher by $\sim$0.5dex for [Fe/H]$>$$-$0.5 dex.}
\label{fig-nafe2}
\end{figure}

Figure~\ref{fig-nafe2} is similar to Figure~\ref{fig-nafe1}, but also includes the
C10 Sgr and M54 data points, and a larger scale.
A striking feature in Figure~\ref{fig-nafe2} is the high [Na/Fe] ratios for the most
metal-rich Sgr stars in the C10 sample; this is inconsistent with all other Sgr 
studies to date.   We believe that these high C10 [Na/Fe] ratios are probably erroneous,
perhaps resulting from blended and saturated Na~I lines 
measured with relatively low resolution spectra. In our spectra the 5682
and 5688\,\AA\ lines, used by C10, are too strong and blended to give reliable results; 
consequently, in this work we derived Na abundances from the neutral lines at 6154 and 6161\,\AA .

We note that C10 applied the non-LTE corrections of Gratton et al. (1999) 
to their abundances.
Gratton et al. (1999) suggest typical Na
non-LTE abundance corrections at $+$0.10 to $+$0.17 dex (for [Fe/H]= $-$0.5 and $-$0.1);
this falls far short of the $\sim$0.5 dex upward shift required to bring SM02, B00, and our
results close to the C10 values.  Thus, C10's non-LTE corrections cannot explain their
relatively high [Na/Fe] ratios, as seen in Figure~{\ref{fig-nafe2}}.  
Calculations by Lind et al. (2011) found non-LTE corrections for Na~I lines in cool giants 
near 0.1--0.2 dex, similar to Gratton et al. (1999).  However, Lind et al. (2011) did not
confirm the trend to large non-LTE corrections, up to $+$0.5 dex, for the lowest
gravity stars claimed by Gratton et al. (1999).  
A check on the non-LTE abundance effect on Na in cool red giant stars comes from a comparison
of LTE [Na/Fe] ratios for K giants and FGK dwarf stars in the Galactic disk, performed by
Fulbright et al. (2007).  The excellent agreement between the dwarf and RGB star 
[Na/Fe] versus [Fe/H] trends indicated that non-LTE effects must be similar and likely
small.

The main nucleosynthesis source of sodium is thought to be carbon burning, after core
carbon ignition in massive stars that ultimately end as SNII (e.g., Woosley \& Weaver 1995).
However, roughly 10\% of the $^{23}$Na is produced by hot-bottom proton burning of
hydrogen-rich envelope material by the Ne-Na cycle (Woosley \& Weaver 1995). 

McWilliam \& Smecker-Hane (2005a) and SM02 pointed-out that the Na and Al deficiencies 
are consistent with a paucity of material ejected from core-collapse supernovae and/or a
low SNII/SNIa ratio, similar to the argument for the low [$\alpha$/Fe] ratios in
dwarf galaxies.  

The extensive detailed abundance study of M54 by C10 found oxygen deficiencies and sodium
enhancements for a large fraction of the cluster stars.  This Na--O anti-correlation is seen
in most Galactic globular clusters (e.g., Carretta et al. 2009) and is evidence that the cluster 
was polluted by proton-burning products during its formation.  As shown by C10, M54 stars exhibit
a tight correlation in the [Na/Fe] versus [O/Fe] plane, presumably due to dilution of the proton
burning products with unprocessed material.  We note that our three stars, and those of S07, lie far
from the locus of points in C10's plot of [Na/Fe] versus [O/Fe] for M54, consistent with the idea
that our stars are members of Sgr, rather than M54, showing no detectable signature of
pollution by proton-burning products.

\begin{figure}[h]
\centering
\includegraphics[width=8.0cm]{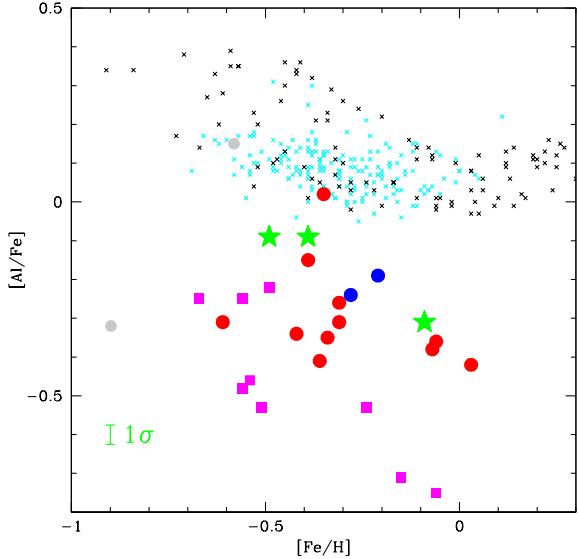}
\caption{[Al/Fe] for our program stars (filled green stars) compared to the Galactic thin disk
measurements of Bensby et al. (2005; black crosses) and Reddy et al. (2003; cyan crosses).
Sgr values from other studies are the same as in Figure~\ref{fig-alphas}.
Of our stars, only \#247 shares the $\sim$0.3 dex deficiency found by SM02 for Sgr, while stars
\#242 and \#266 lie between the MW disk and other reported Sgr 
[Al/Fe] ratios.  S07 give the lowest [Al/Fe] values.  Generally, the S07 [X/Fe] ratios appear lower than
the SM02 and C10 studies of Sgr.}
\label{fig-alfe}
\end{figure}

In Figure~\ref{fig-alfe} we compare [Al/Fe] measured here with results from previous studies
of Sgr and the Galactic thin disk; symbols are the same as in Figures~\ref{fig-alphas}, 
\ref{fig-nafe1}, and \ref{fig-nafe2}.  The median [Al/Fe] across all studies shown in 
Figure~\ref{fig-alfe} is
near $-$0.3 dex.  However, results for stars 242 and 266 in this work are slightly higher,
both at $-$0.09 dex.
Our star 247, at [Al/Fe]=$-$0.31 is similar to SM02, but 
about 0.2 dex higher than the [Al/Fe] ratios found by S07.  
Carretta et al. (2010) measured [Al/Fe] for only two Sgr stars, at $+$0.15 
and $-$0.32.  In general, our [Al/Fe] ratios confirms the Al deficiencies found in all 
other studies of Sgr; however, it remains to be determined whether the small differences
between studies are real.  Similar, low-[Al/Fe], ratios have also been identified in the 
Sculptor dwarf galaxy by Geisler et al. (2005).  Al abundances have not often been measured
for stars in other local group dwarf galaxies, perhaps because the lines were not detected.

As with other low-ionization neutral species, it seems possible that non-LTE effects might
significantly affect the computed LTE Al abundances.  Some non-LTE studies for Al lines
have been performed (e.g., Gehren et al. 2006; Andrievsky et al. 2008); however, these
have focussed on metal-poor stars, much more metal-poor than our Sgr stars, and/or
stars whose temperatures are $\sim$2000K hotter than the cool red giants studied here.
An eye-ball extrapolation of the Andrievsky et al. (2008) non-LTE corrections to the
temperatures and [Fe/H] values of our stars suggests corrections $\sim$0.2--0.3 dex,
with a considerable uncertainty.  The Gehren et al. (2006) non-LTE calculations suggest
corrections of $\sim$0.1--0.3 dex for our stars, although no extrapolation in gravity
parameter is possible.

As with Na, Fulbright et al. (2007) compared LTE [Al/Fe] versus [Fe/H] trends from
RGB and dwarf stars in the Galactic disk.  They found a systematic offset, such that
the RGB star [Al/Fe] ratios needed to be reduced by 0.08 dex in order to come into agreement
with the dwarfs.  This 0.08 dex offset likely reflects the difference in non-LTE correction
between the dwarfs and giants.  
This suggests that our RGB [Al/Fe/] ratios probably need to be revised down by 0.08 dex, in
order to compare to the Galactic disk dwarf trend.  This would serve to increase the
apparent Al deficiency in Sgr, with the non-LTE corrected [Al/Fe] near $-$0.4 dex.


The [Al/Mg] ratios of our Sgr stars show no sign of the anti-correlation seen
in globular cluster stars affected by proton burning products. Similarly, our [Al/Mg] ratios
differ significantly from the locus of globular cluster stars in M54, whose composition 
does exhibit signs of proton burning in the study of C10.  Thus, there appears to be no
evidence of proton-burning products in the atmospheres of our Sgr stars, based on both the
[Al/Mg] and [Na/O] ratios.

Calculations by Woosley \& Weaver (1995) showed that most Al is produced in hydrostatic
carbon and neon burning; thus, like Na, Al is produced mostly by massive stars that end 
as SNII.  However, some Al is produced in the envelopes of intermediate mass AGB stars 
stars by proton burning in the Mg--Al cycle (e.g., Karakas \& Lattanzio 2003).  Since
Sgr stars show no evidence of contamination by such proton burning products,
we conclude that the observed Al deficiencies result from a paucity of
SNII material in this galaxy compared to the MW disk.


\subsection{Iron-Peak Elements}

In addition to iron we have also measured LTE abundances for the iron-peak elements
V, Mn and Cu.  All of these elements have odd-numbers of protons and consequently
possess strong hyperfine splitting, which affects line formation through de-saturation.
Abundances were computed using the measured EWs and the hfs line lists shown in
Table~\ref{tab-hfs}.

\subsubsection{Vanadium}

The average [V/Fe] of our 3 stars is $-$0.02$\pm$0.07 dex, completely consistent with
the solar value.  This is at odds with the results from S07, who found general vanadium 
deficiencies in Sgr, with an average [V/Fe]=$-$0.40$\pm$0.05 dex.  However, 
from 14 Sgr stars Smecker-Hane \& McWilliam (unpublished) found the average
[V/Fe]=0.00 with 1$\sigma$ scatter of 0.13 dex, in good agreement with this work.
C10 did not measure vanadium abundances.  We suspect that the S07 V deficiencies are
spurious, possibly due to adopted low T$_{\rm eff}$ values.  Low temperatures may
also account for the lower [$\alpha$/Fe] values found by S07 than other studies.
However, the S07 results may indicate that V is more deficient farther from the
Sgr nucleus.

On the other hand, expectations from chemical evolution models and supernova
nucleosynthesis (e.g., Timmes, Woosley, \& Weaver 1995; Woosley \& Weaver 1995; Arnett 1971)
predict sub-solar [V/Fe] ratios at low [Fe/H], which has not yet been found
in MW stars.

\subsubsection{Manganese}

\begin{figure}[b]
\centering
\includegraphics[width=8.0cm]{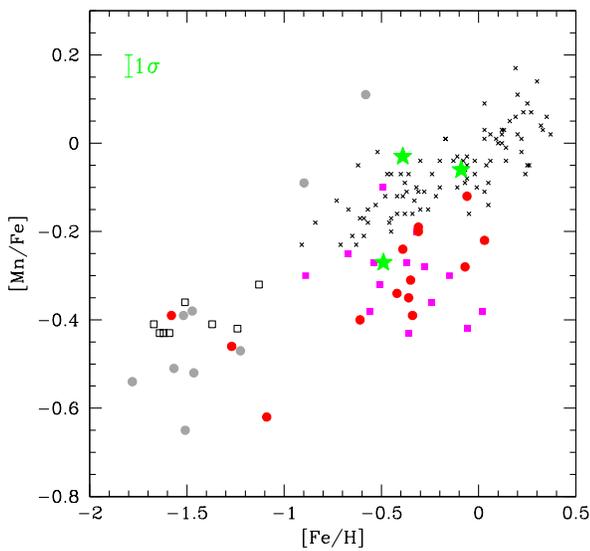}
\caption{[Mn/Fe] for our program stars (filled green stars) compared to Sgr 
results from SM02 (filled red circles)  and S07 (magenta squares).  Two Sgr stars
and 7 M54 stars from C10 are indicated by grey filled circles.  Also shown are solar 
neighborhood points from Feltzing et al. (2007; crosses) and Sobeck et al. (2006; black open squares).
Our new results appear to be in better agreement with the Milky Way trend,
rather than previous Mn measurements for Sgr.}
\label{fig-mnfe}
\end{figure}

The study of the [Mn/Fe] trend in Sgr, the MW disk, and the sun is
filled with contradictory conclusions; regrettably, our results add to this confusion.
The one thing everyone agrees upon is that [Mn/Fe] increases from 
approximately $-$0.4 dex at [Fe/H] typical of the MW halo
to $+$0.1 to $+$0.2 dex for disk stars with super-solar [Fe/H].  The Mn deficiency in
metal-poor MW halo stars was first noted by Wallerstein (1962).

In Figure~\ref{fig-mnfe} we compare our [Mn/Fe] ratios with the MW disk and halo results from
Feltzing et al. (2007, black crosses) and Sobeck et al. (2006, black open squares) respectively.
Our three Sgr stars agree with the trend of these MW results.

It is notable that the S07 and McWilliam, Rich \& Smecker-Hane (2003) [Mn/Fe]
trends for Sgr lie $\sim$0.2 dex below the MW trend, and thus lower than
the results for the three stars in this paper; although, star 242 in this study is
reasonably consistent with the previous studies of Mn in Sgr.  Caretta et al. (2010) 
measured [Mn/Fe] for only 2 Sgr stars, but their values lie well above the trend in 
the MW disk
and much higher than all other Sgr studies.  Carretta et al. (2010) also measured [Mn/Fe] for
7 stars in M54, and found values $\sim$0.1 dex lower than the Galactic globular cluster values
of Sobeck et al. (2006).  If we take this to indicate that a $+$0.1 dex correction is required 
for the C10 [Mn/Fe] values, this would increase the discrepancy between their
two Sgr stars and all other studies.  At the very least, it seems that the [Mn/Fe] values for
the two Sgr stars in C10 are so anomalous that they are suspect.


While the [Mn/Fe] results for Sgr stars in this work are higher than other 
studies, we are wary about preferring one result over another.  An analysis of Mn in the
sun, by Bergemann \& Gehren (2007), showed that while non-LTE corrections to the solar
Mn abundance were of order 0.05 dex, the laboratory oscillator strengths for the transitions show
larger than expected discrepancies between studies, of order 0.1 dex.  Bergemann \& Gehren (2007)
also derive systematically low Mn abundances for Mn~I lines arising from levels with
excitation potentials of 2--3 eV, similar to problems found for solar Fe~I lines
(e.g., Blackwell et al. 1982).  In addition, there are long-standing differences between solar
photospheric abundances from Mn~I lines and the meteoritic value, sometimes by as much as 0.3 dex.
Bergemann \& Gehren (2007) confirm this lacuna and conclude that non-LTE and log~gf value problems
alone cannot account for such deviations, and they suggest that 3D hydrodynamical calculations may
be required to understand the discrepancies.
These problems suggest that the most robust way to estimate [Mn/Fe] values is with
line-by-line differential abundance analysis, which is the method we have employed in this work.
It is clear that further study is required in order to determine whether the trend of
[Mn/Fe] versus [Fe/H] is lower in Sgr than the MW disk.

The [Mn/Fe] differences may be due to the
absolute abundance technique employed by SM02 and McWilliam et al. (2003), and the paucity of
photometric and reddening information available at the time, which made it difficult to
constrain the atmosphere parameters of their stars.  In contrast, for the current work we had
access to extensive optical and infrared photometric data, and we have employed our line-by-line
differential abundance method, relative to Arcturus, which we believe is superior to the method
used by SM02.  Furthermore, the [Mn/Fe] uncertainties of the absolute method were increased by the
discrepancy between published solar photospheric and meteoritic values.  While we believe that the techniques
employed in this work are superior to SM02 and McWilliam et al. (2003) it is still possible that the
[Mn/Fe] discrepancy could be due to the relatively low S/N of the current spectra.

If the final conclusion reached is that the [Mn/Fe] ratios in Sgr are deficient relative
to the MW disk trend, then the conclusion of McWilliam et al. (2003) and Cescutti et al. (2008)
holds: that metal-poor SNIa contributed a significant portion of the iron-peak material to
the more metal-rich Sgr stars.  This could have occurred during leaky-box chemical evolution,
expected of dwarf galaxies.

Recently, Cunha et al. (2010) have found Mn deficiencies in stars belonging to the large
Galactic globular cluster Omega~Cen, the only other place where [Mn/Fe] deficiencies relative
to the MW trend have been claimed.

\subsubsection{Copper}

\begin{figure}[h]
\centering
\includegraphics[width=8.0cm]{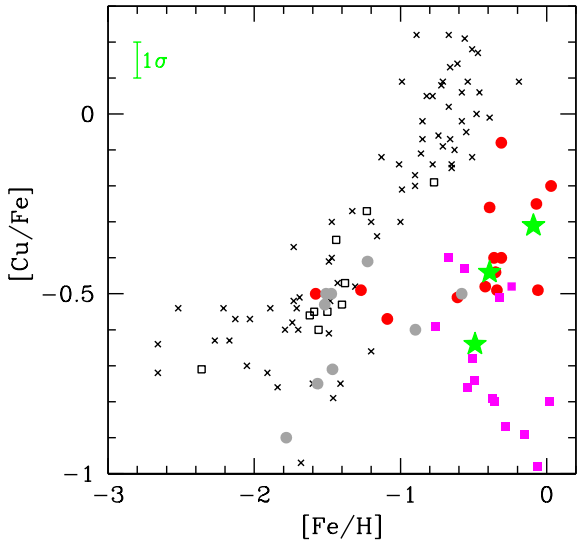}
\caption{[Cu/Fe] for our program stars.  Symbols are the same as in Figure~\ref{fig-alphas}.
Also shown are solar neighborhood points from
Simmerer et al. (2003; black open squares) and Mishenina et al. (2002; crosses).
Clearly, Cu is deficient in Sgr compared to MW stars.}
\label{fig-cufe}
\end{figure}

Figure~\ref{fig-cufe} shows the [Cu/Fe] trend with [Fe/H] for our three stars
compared to other Sgr studies, as well as M54 from C10, and MW stars
from Mishenina et al. (2002) and Simmerer et al. (2003).  Inspection of Figure~\ref{fig-cufe}
shows that our results share the Cu deficiencies seen in all studies of Sgr.
However, our results are most similar to that of McWilliam \& Smecker-Hane (2005b), with $\sim$0.5 dex
[Cu/Fe] deficiency compared to MW stars; this agreement is not surprising given that
the Cu hfs list was the same in the two studies.  Although C10 only measured [Cu/Fe] for two
Sgr stars, the results are in good agreement with this work and McWilliam \& Smecker-Hane (2005b).
On the other hand, S07 found [Cu/Fe] deficiencies that increased with increasing [Fe/H],
such that by solar iron abundance the [Cu/Fe] ratio is near $-$1 dex.  Whether this difference
is real needs to be further investigated.  If the extra Cu-deficiencies in S07 are real it would
suggest chemical inhomogeneity, or a Cu gradient, in Sgr, since the S07 field is
relatively distant from other Sgr studies considered here.  
One unexpected observation is that the C10 [Cu/Fe] ratios for M54 stars show
a large range, roughly 0.6 dex; this might simply reflect large measurement uncertainties
due to the relatively low-resolution spectra employed by C10.

Similar [Cu/Fe] deficiencies to those found for Sgr have been measured in the massive 
Galactic globular cluster $\omega$ Cen by Cunha et al. (2002) and Pancino et al. (2002); notably,
$\omega$ Cen also shows Mn deficiencies, similar to Sgr.  Pomp\'eia et al. (2008) found
[Cu/Fe] deficiencies even for the highest [Fe/H] stars in the LMC; this adds to the chemical similarities
of the LMC and Sgr, which includes sodium and alpha-element deficiencies and s-process enhancements.
Nissen \& Schuster (2011)
have also found a sub-population of MW halo stars showing deficient Cu abundances, in addition to 
low [$\alpha$/Fe], [Na/Fe] and [Al/Fe] ratios and high [Ba/Y] values; again, these abundance ratios are
similar to those of Sgr.  
We agree with the conclusion of Nissen \& Schuster (2011), that this sub-population reflects the
accretion of late-time dwarf galaxies into the Galaxy.

\begin{figure}[h]
\centering
\includegraphics[width=8.0cm]{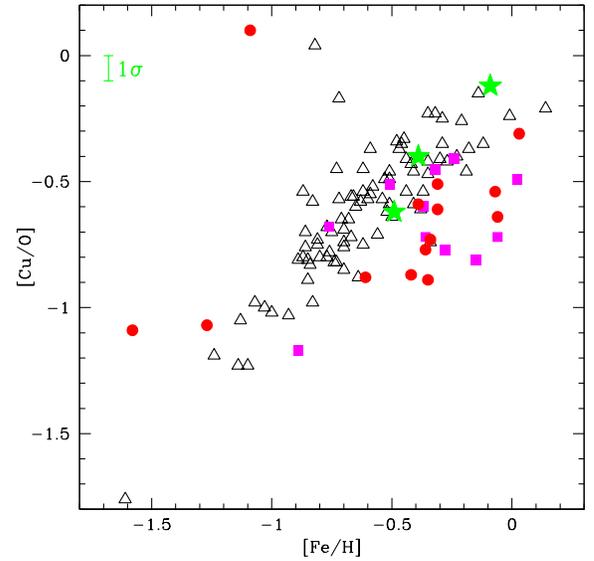}
\caption{[Cu/O] versus [Fe/H] for our program stars and other Sgr studies (symbols the same as in
Figure~\ref{fig-alphas}) compared to the thick disk results of Reddy et al. (2006; open black triangles).
The similar metallicity-dependent [Cu/O] trend for the thick disk and Sgr is
consistent with the idea that Cu is produced by the progenitors of Type~II SNe, in agreement with
Bisterzo et al. (2004) and many others.  However, S07 and SM02 results show [Cu/O] ratios lower
by$\sim$0.2 dex, although they also show an increasing trend with [Fe/H].}
\label{fig-cuo}
\end{figure}

Copper is thought to be predominantly produced in the hydrostatic He and C-burning phases of 
massive stars (which ultimately become SNII) via weak s-process neutron capture, driven by
$^{22}$Ne($\alpha$,n)$^{25}$Mg.  This idea has been developed from calculations of the s-process
in massive stars, including papers by Prantzos et al. (1990), Raiteri et al. (1993), The et al. (2000),
Bisterzo et al. (2004), Chieffi \& Limongi (2006), Pignatari et al. (2010),  and 
Pumo et al. (2010, 2012).  These papers indicate that the yield of copper increases with
increasing metallicity, as expected
from the metallicity-dependence of the s-process, but also with the mass of the massive star, presumably
due to the size of the He and C-burning regions.  However, significant complications
arise in computing the Cu yields, due to such effects as nuclear reaction rates, convective overshoot,
mass-loss and fall-back; a number of the aforementioned papers discuss these difficulties.
In addition, a minor component of the Cu is thought to be produced during explosive nucleosynthesis
(e.g. Woosley \& Weaver 1995).

In Figure~\ref{fig-cuo} we show that the trend of [Cu/O] versus
[Fe/H] in the MW thick disk and Sgr closely follow each other; this work
supports the idea that Cu is mainly produced by massive stars that end as core-collapse SNe.
However, the SM02 and S07 [Cu/O] ratios lie slightly below the thick disk trend.

These conclusions about
Cu production suggest that environments with a paucity of ejecta from massive stars should show
Cu deficiencies.  Given that massive stars are thought to be the major source of 
$\alpha$-elements and Na and Al, the deficient Cu abundances should be accompanied by 
low $\alpha$, Na and Al abundances.  These abundance patterns are, indeed, found in this 
work and in other studies of Sgr.  Thus, the low [Cu/Fe] ratios suggest a low SNII/SNIa
ratio.  Low SNII/SNIa ratios may follow some time after a burst of star formation,
due to the time delay between SNII and SNIa, or can occur as a result of an IMF deficient 
in high-mass stars (either top-light or from a steep IMF slope).

\subsection{Hydrostatic and Explosive Elements}

At this point we have mentioned a number of elements whose synthesis in massive stars
is dominated by either hydrostatic helium, carbon, or neon burning phases (e.g., O, Mg, Na, Al 
and Cu), or in the SNII explosion event (Si, Ca, Ti, and Fe).  The yield of the hydrostatic elements
increases with stellar mass; for Al, Na and Cu the yields are also thought to be
metallicity-dependent (e.g., Arnett 1971; Prantzos et al. 1990; Woosley \& Weaver 1995), despite the observed
flat trend of [Na/Fe] in the Galactic disk.  

The explosive elements are thought to be produced in both SNII and SNIa events, with lower
[X/Fe] ratios from SNIa than for SNII events.  However, the exact SNIa [Si/Fe], [Ca/Fe] and
[Ti/Fe] yield ratios are not known, but lie somewhere below the solar value.  Production
of the hydrostatic elements by SNIa is thought to occur, but with negligibly small [X/Fe] ratios
(e.g., Nomoto et al. 1984; Maeda et al. 2010).  In this section we compare the abundances of
the hydrostatic and explosive element families in Sgr with the MW thick disk.

To understand how Sgr evolved we wish to compare our measured element
abundance ratios to some standard population.  We choose the MW thick disk as
our standard reference because the thick disk mean [Fe/H], near $-$0.6 dex, and [Fe/H] range,
approximately from $-$2.0 to 0.0 dex, is very similar to Sgr,
because the thick disk stellar ages cover a range of $\sim$5 Gyr (Reddy et al. 2006), 
similar to the age difference between the two Sgr populations studied by Siegel et al. (2007),
and because the thick disk composition is well measured.  
Thus, the metallicities and timescales
are similar for these two systems, so chemical composition differences are 
less likely to be due to metallicity or timescale-dependent parameters.

\begin{figure*}[ht]
\centering
\includegraphics[width=19.0cm]{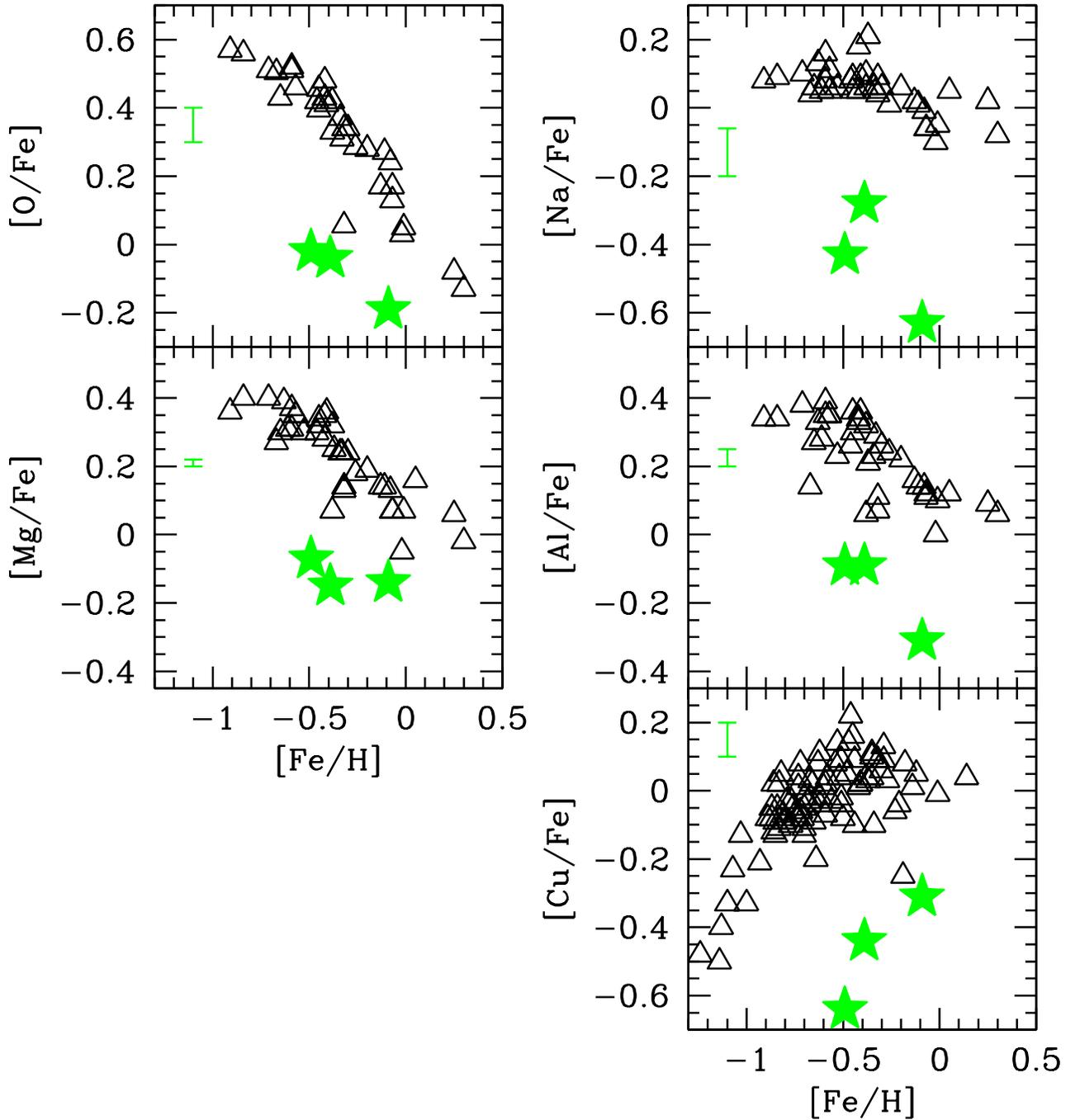}
\caption{A comparison of the hydrostatic alpha elements, O and Mg, Na, Al, and Cu in 
Sgr (filled green stars) with the thick disk results of Bensby et al. 
(2005; open black triangles).  The figures show that elements produced in the hydrostatic
burning phases of of Type~II SNe progenitors are deficient, relative to iron, in all cases
by more than a factor of two.}
\label{fig-omgnaalcu}
\end{figure*}

\begin{figure*}[ht]
\centering
\includegraphics[width=17.0cm]{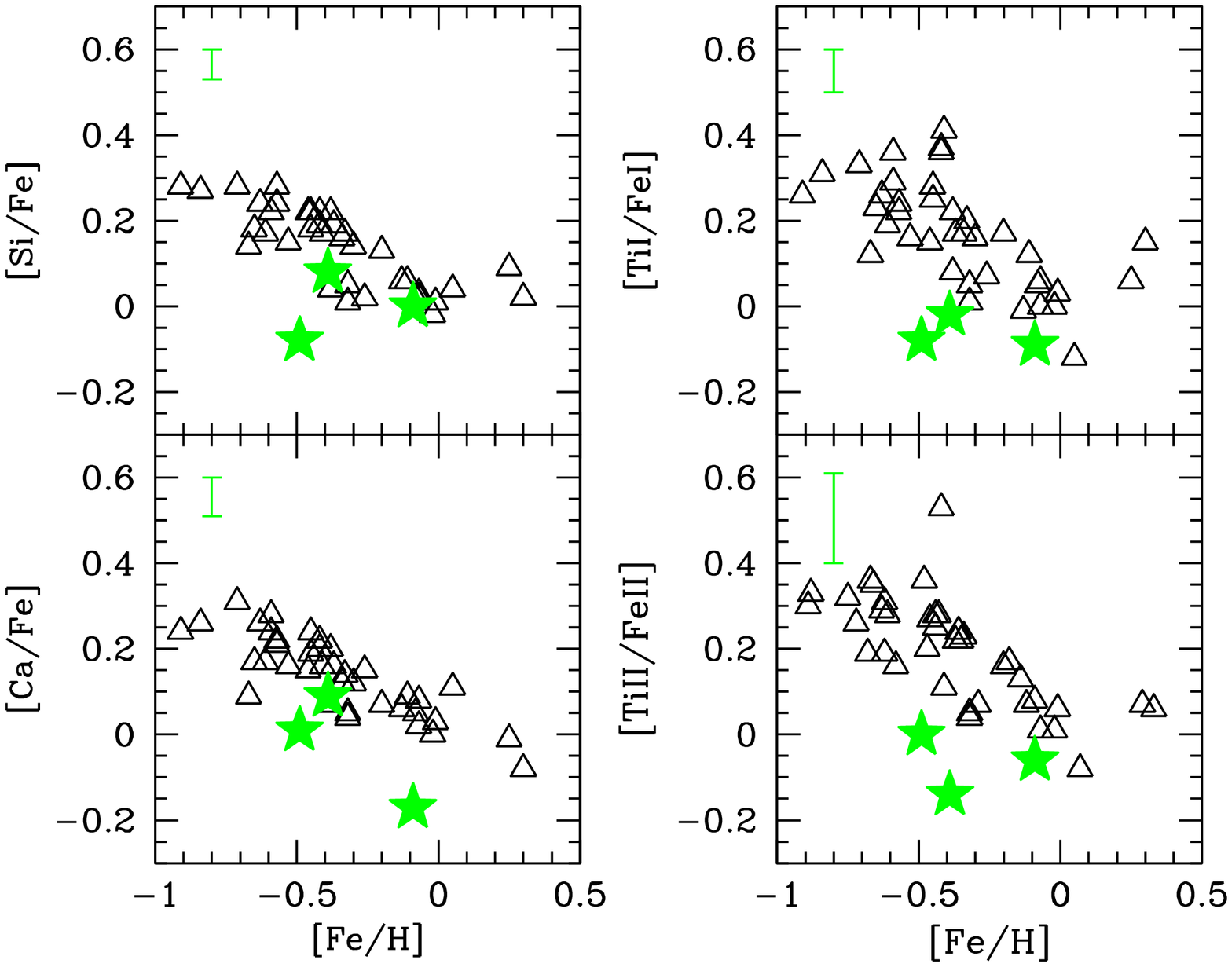}
\caption{Comparison of the explosive alpha elements, Si, Ca and Ti, 
in Sgr (filled green stars) with the thick disk results of Bensby et al. 
(2005; open black triangles).  
The figures show that explosive nucleosynthesis products of Type~II
SNe are deficient, relative to iron, by $-$0.17 dex, which is significantly less than the
deficiency for the hydrostatic elements, at $-$0.43 dex.}
\label{fig-sicati}
\end{figure*}

In Figure~\ref{fig-omgnaalcu} we compare our Sgr hydrostatic element ratio measurements for
[O/Fe], [Mg/Fe], [Na/Fe], [Al/Fe] and [Cu/Fe] with the thick disk results of Bensby et al. (2005).  
For the explosively produced $\alpha$-elements, we compare our Sgr [Si/Fe], [Ca/Fe], [Ti~I/Fe]
and [Ti~II/Fe] ratios to Bensby's thick disk trends in Figure~\ref{fig-sicati}.
A glance at Figures~\ref{fig-omgnaalcu} and \ref{fig-sicati} shows that the Sgr hydrostatic 
element [X/Fe] ratios lie much further below the thick disk trend than the ratios for the 
explosively produced elements.

In Table~\ref{tab-diffab} we list the $\Delta$[X/Fe] abundance shifts,
which would move the observed Sgr [X/Fe] ratios to the thick disk trends;
thus, the shift indicates the change in X required to transform
the thick disk abundance ratios into the measured Sgr values.

Clearly, in Figures~\ref{fig-omgnaalcu} and \ref{fig-sicati} the addition of some
amount of extra Fe to the MW trends can be used to reproduce the measured Sgr [X/Fe]
and [Fe/H] ratios for each element, X; however, the amount of Fe required is different
for each element: 0.7 dex is required for O, 0.5 dex for Na, Al and Mg, 0.2 dex for Ca and Si,
0.3 dex for Ti, 0.4 dex for Cu, and 0.0 dex for r-process Eu.  Thus, it is not possible to 
reproduce the Sgr composition by adding a single quantity of Fe to the MW disk ratios, even
for pure SNII elements (such as O, Mg, Al, Na and Cu); therefore, we do not consider this
possibility further. 

\begin{deluxetable}{lc}
\tabletypesize{\scriptsize}
\tablecaption{Sgr--Thick Disk Element Differences}
\tablewidth{0pt}
\tablehead{\\
\colhead{Species} &
\colhead{$\Delta$[X/Fe] } \cr
 & (dex) \cr }
\cline{1-2}
\cr
\startdata
\cr
[O~I] & $-$0.43 $\pm$ 0.03 \cr 
Na~I  & $-$0.50 $\pm$ 0.09 \cr 
Mg~I  & $-$0.34 $\pm$ 0.05 \cr 
Al~I  & $-$0.39 $\pm$ 0.04 \cr 
Si~I  & $-$0.14 $\pm$ 0.06 \cr 
Ca~I  & $-$0.16 $\pm$ 0.05 \cr 
Ti~I  & $-$0.20 $\pm$ 0.07 \cr 
Ti~II & $-$0.21 $\pm$ 0.08 \cr 
Cu~I  & $-$0.50 $\pm$ 0.11 \cr 
\enddata
\label{tab-diffab}
\end{deluxetable}

Table~\ref{tab-diffab} demonstrates that the hydrostatic elements all possess rather large
deficiencies, $\Delta$[X/Fe], relative to the Galactic thick disk trend.  
Within this group of hydrostatic elements, the individual $\Delta$[X/Fe] values show a dispersion 
that may be real, for example $\Delta$[Mg/Fe]=$-$0.34 dex while $\Delta$[Na/Fe]=$-$0.50 dex;
however, the group is reasonably represented by the mean $\Delta$[X/Fe]=$-$0.43$\pm$0.03 dex.
Similar alpha-element deficiencies, have been found in previous abundance studies of
dwarf galaxies (e.g., Shetrone et al. 2001, 2003; Geisler et al. 2005) as well as in
sub-populations of the Galactic halo (e.g., Brown et al. 1997; Nissen \& Schuster 1997);
also see Venn et al. (2004).

In Table~\ref{tab-diffab} the explosive elements show smaller deficiencies for [Si/Fe], 
[Ca/Fe] and [Ti/Fe], relative to the thick disk, at $-$0.14, $-$0.16 and $-$0.21 dex 
respectively, with a mean $\Delta$[X/Fe]=$-$0.17$\pm$0.03 dex.   Thus, the 
explosive element [X/Fe] ratios, relative to the thick disk, are 0.25 dex higher
than for the hydrostatic elements.

\begin{figure}[ht]
\centering
\includegraphics[width=8.0cm]{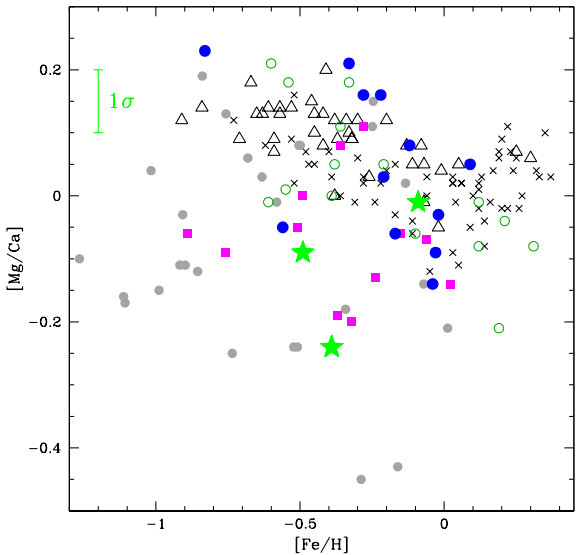}
\caption{The trend of [Mg/Ca] versus [Fe/H] in the thin disk (crosses Bensby et al. 2005),
thick disk (black open triangles: Bensby et al. 2005) and Sgr (symbols the same as Figure~\ref{fig-alphas}
except B00 and B04 are both represented by filled blue circles).  The Sgr trend for C10 and S07
is $\sim$0.2 dex lower than the Galactic disk, while Sgr B00/B04 points are similar to the disk;
our average [Mg/Ca] is lower than the MW disk by $\sim$0.1 dex.  Both Sgr and the Galactic disk
show a decline in [Mg/Ca] with increasing [Fe/H].}
\label{fig-mgcafeh}
\end{figure}

The unusually low abundance ratio of hydrostatic to explosive elements is also seen in the 
[Mg/Ca] ratios from previous Sgr studies, as shown in Figure~\ref{fig-mgcafeh}.
The Sgr results of C10 and S07 are, on average, deficient in [Mg/Ca], relative to the MW disks,
by $-$0.19 dex, in agreement with our result.  However, the [Mg/Ca] ratios given by B00 and B04 
are similar to the MW disk trend.  These details notwithstanding, all studies show a steady decline 
in [Mg/Ca] with increasing [Fe/H], by 0.4 dex/dex for Sgr and 0.2 dex/dex for the MW disk.  

The $\Delta$[X/Fe] differences between hydrostatic and explosive elements suggests a relative paucity
of nucleosynthesis products from the most massive SNII events, a conclusion that is bolstered by
our inclusion of Na, Al and Cu as hydrostatic elements.  These deficient hydrostatic/explosive
abundance ratios can be explained by at least two scenarios: enrichment by an IMF deficient in the
most massive SNII progenitors, or from nucleosynthesis with excess SNIa, perhaps due to long-lived 
SNIa progenitors that contribute metals over an extended period.

Support for both mechanisms can be found in the literature: Weidner \& Kroupa (2005) and 
Kroupa et al. (2011) predicted a steeper integrated-galactic IMF (IGIMF) for dwarf galaxies (like
Sgr).  They argued that, with less total gas mass than the MW, dwarf galaxies lack the most massive
molecular clouds and so are less efficient at producing the most massive stars.  Kroupa et al. (2011)
also predicted an IGIMF slope which steepened with increasing metallicity.  Likewise, Oey (2011)
showed that the IMF slope steepens when there is insufficient gas to make the largest molecular clouds.
Our Sgr low hydrostatic/explosive abundance ratios, including [Mg/Ca], suggest an 
IMF deficient in the most massive stars, because the yield of hydrostatic elements increases with
increasing SNII progenitor mass.

On the other hand, the SNIa time-delay scenario of Tinsley (1979) has long been invoked to 
explain the decline in [O/Fe], and other $\alpha$-elements, with [Fe/H] in the MW disks.  Indeed,
Matteucci \& Brocato (1990) predicted deficient [O/Fe] ratios in the LMC due to its presumed
low SFR compared to the MW.  Our low hydrostatic/explosive element ratios might also be 
understood in this scenario, since SNIa produce Si, Ca and Ti but not O, Na, Mg, Al and Cu in
significant quantities.  A particular difficulty is that Fe and the explosive alphas can be
produced by both low-mass SNII and by SNIa, so without good constraints on the element yields it
is not easy to disentangle the relative role of these two nucleosynthesis sources.
Thus, the decline of [Mg/Ca] in the MW disks and Sgr could be explained by a metallicity-dependent
IMF over the range from [Fe/H]=$-$1 to 0, or by an increasing nucleosynthetic contribution
of Ca from SNIa events over the formation times of the disk and Sgr.

Tolstoy et al. (2003) claimed that the low [$\alpha$/Fe] ratios seen in dwarf galaxies were evidence
of steep IMFs, with reduced contributions from massive stars; however, their argument was specious as
it omitted other scenarios.  In a subsequent paper the same group (Venn et al. 2004) asserted that low
[$\alpha$/Fe] ratios could occur without affecting the IMF, by the addition of material from SNIa.

We note that our [O/Mg] ratios in Sgr, correlated with O/hydrostatic ratios, lie precisely on the
declining [O/Mg] trend with [Mg/H] for the MW bulge and disks, noted by McWilliam et al. (2008).  
The decline in [O/Mg] is thought to be due to a quenching of oxygen yields from massive stars as
 metallicity-dependent winds strip their outer layers (McWilliam \& Rich 2004; McWilliam et al. 2008; 
Cescutti et al. 2009).  From [Fe/H]=$-$1 to the solar value, at least $\sim$0.2 dex of the decline in [O/Fe]
must be due to metallicity-dependent stellar wind effects.

\subsection{The s-Process Elements}


In this work we employ La~II lines to indicate abundances for elements in the second
s-process peak.  The La~II lines are strong enough for reliable EW measurement and the
hyperfine constants are well measured.  Thanks to a large nuclear spin, $I=7/2$, the 
hfs splittings are so significant that the La~II lines remain on the linear part of 
the curve of growth to quite large EWs; this greatly enhances the  accuracy of our abundance
measurements.  However, we are not able
to measure reliable barium abundances, because even the weakest Ba~II line, at 5853\AA , is too
strongly saturated (EW$>$200m\AA ), even for the most metal-poor star in our sample.
This is a consequence of the large s-process enhancements in Sgr.

\begin{figure}[h]
\centering
\includegraphics[width=8.0cm]{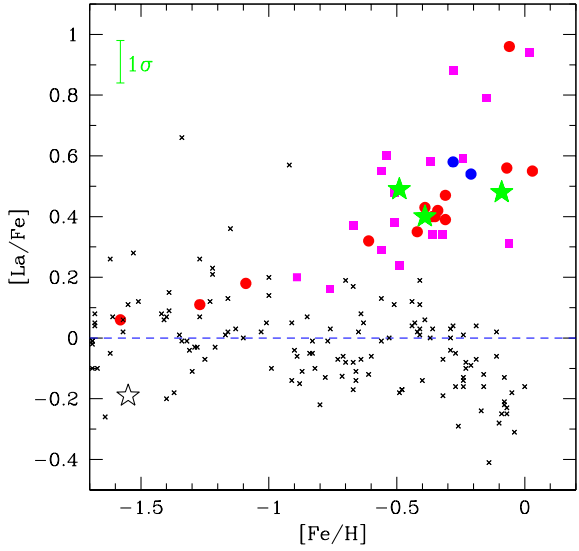}
\caption{[La/Fe] in Sgr stars (see Figure~\ref{fig-alphas} for symbol key).
Also shown are solar neighborhood points from Simmerer et al. (2004; crosses) and
M54 from Brown \& Wallerstein (1999; black open star).}
\label{fig-lafe}
\end{figure}

Figure~\ref{fig-lafe} shows [La/Fe] versus [Fe/H] found here, compared to the Sgr 
results of SM02, S07, B00, and the solar neighborhood Galactic 
stars of Simmerer et al. (2004) .  Our [La/Fe] ratios agree well with all previous studies
of Sgr, showing enhancements of $\sim$0.5 dex.  This chemical signature further strengthens 
the conclusion
that these stars are, indeed, members of Sgr.  Clearly, the stars of Sgr
are significantly enhanced in La compared to the solar neighborhood, which suggests s-process
enrichment.  This is seen in stars of several other nearby dwarf spheroidal galaxies 
(e.g., Shetrone et al.  2001, 2003; Pancino et al. 2008; Geisler et al. 2005;
Letarte et al. 2010).
It is interesting that both S07 and SM02 both find a single Sgr star with [La/Fe]$\sim$1 dex,
near solar metallicity, significantly higher than the trend of [La/Fe] with [Fe/H].
It is not clear whether this is due to a narrow spike in [La/Fe] or reflects inhomogeneity.
At this point we should note that a study of Sgr M giants, by Chou et al. (2010), found
sub-solar [La/Fe] ratios, near $-$0.2 dex, more than $\sim$0.5 dex lower than the [La/Fe]
ratios found here, and by SM02 and S07; the Chou et al. (2010) results also showed a peak-to-peak
scatter of $\sim$1 dex near solar [Fe/H].  Given the difficulty of measuring [La/Fe] in M giants,
the scatter, and the discordant nature of the Chou et al. (2010) results, we choose
not to use them for further comparison.

We note that in Figure~\ref{fig-lafe} the solar neighborhood points of
Simmerer et al. (2004) show
a small decline in [La/Fe] above [Fe/H]$\sim$$-$0.4 dex.  We believe this trend
to be real, and results from the metallicity-dependent decline in 
the production of the heavy s-process elements (e.g., Gallino et al. 1998; Busso et al. 1999).

\begin{figure}[h]
\centering
\includegraphics[width=8.0cm]{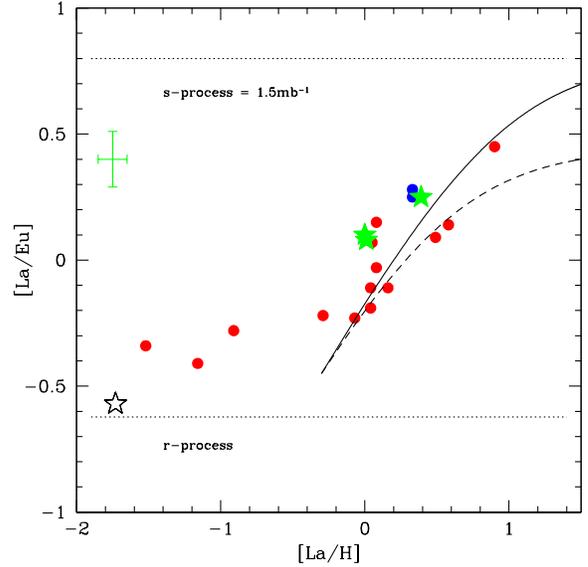}
\caption{[La/Eu] for our program stars (filled green stars) compared to Sgr
following SM02 (filled red circles), two points from B00 (filled blue circles) and
M54 by Brown et al. (1999; open black star).  The horizontal dotted lines show pure
s-process and r-process [La/Eu] ratios. The solid line shows a dilution locus, resulting from
the addition of pure s-process material, while the dashed line shows the dilution locus
for 95\% s-process plus 5\% r-process added to an r-process dominated starting composition.}
\label{fig-laeulah}
\end{figure}

Following McWilliam \& Smecker-Hane (2005a, henceforth MS05) in Figure~\ref{fig-laeulah} we
compare [La/Eu] versus [La/H] for our three stars with SM02 and B00.  
The [La/Eu] ratio distinguishes between r-process and s-process neutron capture.  The
[La/H] abcissa in the plot, as employed by MS05, allows 
metallicity discrimination without the complication of [Fe/H].  Both La and
Eu are made by neutron-capture processes, but Fe is made by in explosive nucleosynthesis
processes by SNIa and SNII.  Thus, plotting [La/Eu] with [La/H] avoids the added complexity,
and uncertainties, due to the production of Fe, in addition to neutron-capture processes.  
The loci 
in Figure~\ref{fig-laeulah} are dilution curves showing the evolution of the composition
with the addition 
of pure s-process material for the solid line, and 95\% s-process with 5\% r-process
for the dashed line.  We note that the three points from this work follow a 
slope roughly consistent with the addition of pure s-process to an earlier composition, 
although offset by $\sim$0.1 dex higher than the values found by SM02 and MS05.
Whether these abundance differences are real, due
to inhomogeneities in Sgr, or resulted from the different 
measurement techniques is not yet certain; however, the agreement is within the measurement
uncertainties.   The similarity of [La/Eu] versus [La/H] in this work with SM02 and
B00 supports our assertion that our stars are, indeed, members of 
Sgr.  As discussed by MS05, the slope of the locus
in Figure~\ref{fig-laeulah} shows enrichment by essentially pure s-process material, with no 
significant r-process production above [La/H]$\sim$$-$0.4 to $-$0.6 dex (corresponding
to [Fe/H]$\sim$$-$0.6 to $-$0.8 dex respectively).  However, the r-process [Eu/H]$_r$ ratios
increase with [Fe/H] among our small sample, showing that some r-process enrichment occurred
during the evolution of Sgr, but at a much lower level than the s-process.

\subsubsection{Heavy and Light S-process}

As discussed by Gallino et al. (1998) and Busso et al. (1999) the ratio of heavy to light
neutron-capture elements, [hs/ls], produced by the s-process,
is sensitive to metallicity.  For the s-process in low-mass AGB stars the
neutrons are predominantly produced via the $^{13}$C($\alpha$,n)$^{16}$O reaction, 
where the $^{13}$C results from
$^{12}$C(p,$\gamma$)$^{13}$C following ingestion of protons from the envelope.  In 
low-metallicity AGB stars, the roughly constant
number of neutrons released in the thermal pulse are captured by very few iron-peak nuclei
and most of the seed nuclei end-up as heavy s-process elements (e.g. Ba, La, Pb).  However, at
higher metallicity the numerous seed nuclei, on average, capture many fewer neutrons and so produce
more of the light s-process elements (e.g., Sr, Y, Zr) than the heavy s-process elements.

\begin{figure}[h]
\centering
\includegraphics[width=8.0cm]{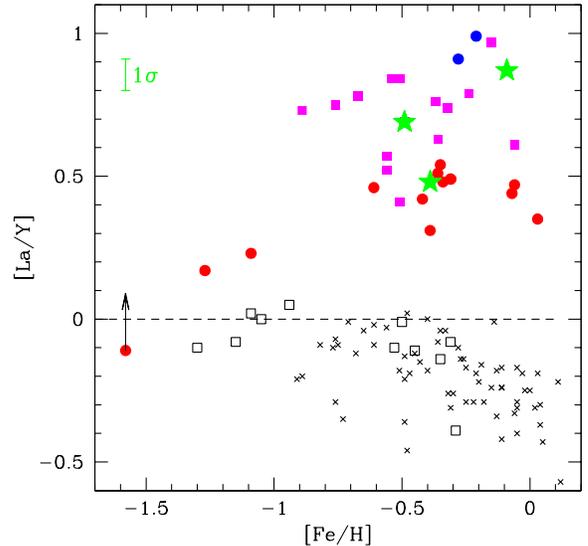}
\caption{[La/Y] for Sgr stars (see Figure~\ref{fig-alphas} for key) and
solar neighborhood stars indicated with black crosses (Simmerer et al. 2002) and
black open squares (Gratton \& Sneden 1994).}
\label{fig-layfeh}
\end{figure}

Figure~\ref{fig-layfeh} shows [La/Y] versus [Fe/H] for our stars, compared to MS05\footnote{MS05 revised
the Y abundances measured by SM02 using $gf$ values from Hannaford et al. (1982)}, S07 and
B00.  The [La/Y] ratios for our stars share the range of [hs/ls]
enhancements seen in nearby dwarf spheroidal galaxies, such as the 0.5--0.8 dex range
found by Shetrone et al. (2001, 2003); see also Letarte et al. (2010), B00 
and SM02.  The [La/Y] enhancements for our Sgr stars indicate s-process nucleosynthesis 
by relatively metal-poor AGB stars [Fe/H]$\sim$$-$0.6 or $\leq$$-$1 dex (Busso et al. 1999), 
but the exact value depends on details of the predicted [hs/ls] curve.
The [Fe/H] values indicated by the measured [La/Y] ratios are lower than 
the [Fe/H] of the stars themselves, particularly the more metal-rich Sgr stars; thus,
neither the stars nor any companion could have produced the observed s-process enrichments. 
This indicates that the nearly ubiquitous s-process enhancements seen in Sgr is
primordial and must have
come from previous generations that enriched the interstellar gas, out of which the current
stars formed.  This 
novel way to produce s-process rich stars (or barium stars), suggested by
SM02, can result from leaky box chemical evolution, where at late times gas from
a large, old, metal-poor, population can dominate the composition of the younger
population of metal-rich stars.

Here we confirm that [La/Y] enhancements are present in Sgr, as found by
B00, SM02/MS05 and S07; our values overlap most with S07 and B00, whilst the SM02/MS05 
values are lower than the current results.


Due to limited wavelength coverage SM02/MS05 had access to 3 to 4 less than optimal Y~II lines,
including the line at 7450\AA , with a poorly known $gf$ value.  
In the present work we have measured [Y/Fe] using 6 Y~II lines for Arcturus and
5 lines in the Sgr stars.  Unlike SM02/MS05, the differential technique used here is
unaffected by poorly known $gf$ values;
however, our spectra have lower S/N than SM02/MS05 and we have only one star with
[Fe/H]$\geq$$-$0.1 dex, while SM02/MS05 had
3 such stars.  The current results, particularly for star~247, combined
with the two points of B00 and the S07 results, indicate an increasing 
trend of [La/Y] with increasing [Fe/H], from [La/Y]$\sim$0.0 dex at [Fe/H]$\sim$$-$0.7 to
[La/Y]$\sim$$+$1 dex by solar [Fe/H].
Such an increase is also seen in the [La/Y] and [Ba/Y] ratios in Fornax, measured by
Letarte et al. (2010).
Because high [La/Y] values are characteristic of nucleosynthesis in low metallicity AGB stars
(Gallino et al. 1998; Busso et al. 1999), it appears that
the most metal-rich Sgr stars formed out of material dominated by ejecta from metal-poor
AGB stars.  Our highest [La/Y] value (for star~247) 
is reproduced by the detailed calculations of Bisterzo et al. (2010), but is higher than
the predictions of Cristallo et al. (2009, 2011) by $\sim$0.25 dex.  A small increase,
in the amount of protons ingested into the intershell region is required, above the standard
treatment (ST), for the Cristallo et al. (2009, 2011) calculations to reproduce the observed
[La/Y] ratio of star~247.  


\begin{figure}[h]
\centering
\includegraphics[width=8.0cm]{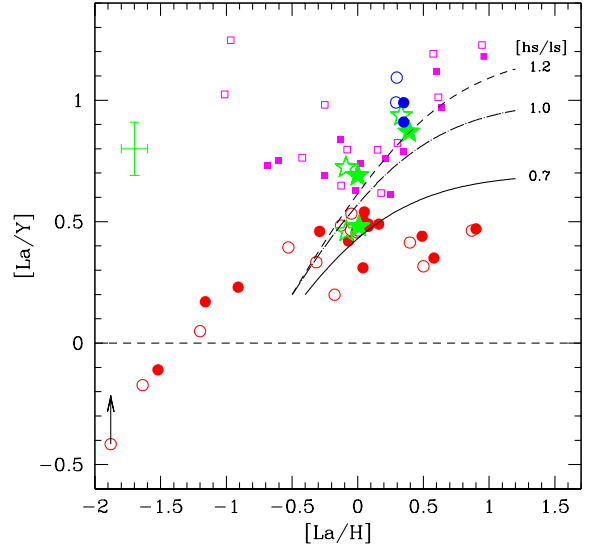}
\caption{[La/Y] versus [La/H] for our Sgr stars (green stars) compared to Sgr results of
SM02 (red circles), B00 (blue circles), and S07 (magenta squares).  Filled symbols
show the measured abundance ratios; open symbols indicate r-process subtracted values.
The solid black line is a dilution curve showing the effect of adding the highest predicted
[La/Y] AGB s-process yields ([La/Y]=$+$0.70 dex) of Cristallo et al. (2011),
to the pre-existing Sgr composition, near [La/H]=$-$0.40 dex ([Fe/H]=$-$0.70).  The dot-dashed
line shows a dilution curve resulting from the addition of [La/Y]=$+$1.00 dex material to 
Sgr gas at [La/H]=$-$0.50 ([Fe/H]=$-$0.76), [La/Y]=$+$0.20, and the short dashed locus 
indicates dilution with [La/Y]=$+$1.2 material.  Except for the SM02  results, the
measurements suggest that the s-process in Sgr produced much higher [La/Y] ratios than 
the AGB predictions of Cristallo et al. (2011).}
\label{fig-laylah}
\end{figure}

Figure~\ref{fig-laylah} shows [La/Y] versus [La/H] for the stars studied in this work, 
compared to Sgr stars studied by MS05/SM02, S07, and B00.  For each source we show the
measured abundance ratios with filled symbols and the r-process subtracted ratios with
open symbols.  The r-process corrections were computed using [Eu/Fe] and [La/Fe] ratios
and adopting r-process ratios for [La/Eu]$_r$=$-$0.58 dex and [La/Y]$_r$=$+$0.57 dex.
These r-process ratios were based on the solar s- and r-process fractions determined by
Bisterzo et al. (2011), Simmerer et al. (2004), Arlandini et al. (1999) and the abundance
ratios in the r-process rich star CS~22892-052 (Sneden et al. 1996).

We used the measured [Eu/Fe] ratios for the r-process corrections to apply to the heavy
element abundances in MS05/SM02, B00 and this work, but for S07 no Eu abundances were
measured.  Consequently, for the S07 points we employed [Fe/H] and the observed
thick disk [Eu/Fe] versus [Fe/H] trend to estimate [Eu/Fe].
That is a reasonable assumption, given the good agreement with the thick disk trend for
r-process [Eu/Fe] ratios measured here and in B00 (see Figure~\ref{fig-eufe_obs}).

Figure~\ref{fig-laylah} shows that the r-process corrections are typically less than
0.1 dex for stars above [La/H]$\sim$$-$0.2 dex and therefore, do not significantly affect
our conclusions.  

It is immediately obvious that the MS05/SM02 results, above [La/H]$\sim$$+$0.3, lie significantly below
the values from 
the three other Sgr studies, suggesting that MS05/SM02 [La/Y] ratios may be in error.
This might reasonably have resulted from the use of blended Y~II lines in the SM02 list, 
or the use of Y~II lines with poorly known $gf$ values in the absolute analysis of SM02.
Such difficulties not withstanding, all Sgr studies show significantly enhanced [La/Y] compared to
the MW disk.

The solid line in Figure~\ref{fig-laylah} represents a dilution curve, starting with the
pre-existing composition at [La/H]=$-$0.40 and [La/Y]=$+$0.20 dex, and adding pure metal-poor
AGB s-process ejecta based on the theoretical [La/Y] yields from Cristallo et al. (2011).
In this case we employ the predictions from their $z=1.0\times 10^{-3}$, 1.5 M$_{\odot}$, model
which happens to give their maximum expected [La/Y] value.  Unfortunately,
this locus severely under-predicts the [La/Y] ratios compared to the majority of Sgr studies, 
including the present work.   If the MS05/SM02 [La/Y] values are disregarded, then the AGB 
s-process [La/Y]
yields must be higher than the Cristallo et al. (2011) predictions.  The dot-dashed line in
Figure~\ref{fig-laylah} shows the dilution locus assuming an intrinsic s-process [La/Y] ratio
of $+$1.00 dex, and starting at [La/H]=$-$0.5 dex (corresponding to [Fe/H]=$-$0.76), while the
short dashed line is the locus for dilution with [La/Y]=$+$1.2 dex. 

These high [La/Y] dilution locii provide a superior comparison to the measured abundance ratios;
however, they are 0.3 and 0.5 dex higher than the maximum predicted AGB s-process [La/Y] yields.
As noted earlier, 
the [La/Y]=$+$1.0 dilution locus is in better agreement with the [Fe/H]=$-$1.0 AGB s-process predictions
of Bisterzo et al. (2010), at [La/Y]=$+$0.9 dex.  However, to obtain [La/Y]=$+$1.00 dex, consistent
with our second dilution curve, would require an increase of the mass of the $^{13}$C pocket
introduced into the intershell region to twice the standard value, or ST*2 in the format of
Bisterzo et al. (2010).

We note that the dilution locii in Figure\ref{fig-laylah} assume that there is no significant
contribution of s-process material from stars more metal-rich than [Fe/H]$\sim$$-$0.5 dex.
If this assumption is incorrect, then higher [Fe/H] AGB material would have been incorporated
into Sgr, with characteristically lower [La/Y].  In that case, to match the [La/Y] ratios
measured in this work, by S07 and B00 would require an even larger increase 
in the mass of the $^{13}$C pocket.  Thus, our dilution curves provide a minimum estimate of
the [La/Y] yield ratio (and $^{13}$C pocket) for AGB stars near [Fe/H]=$-$0.6 dex.

The reasonable fit of the measured [La/Y] ratios to the [La/Y]=$+$1.2 dilution curve
suggests that AGB stars with [Fe/H]$>$$-$0.5 dex did not dominate neutron-capture element
nucleosynthesis in Sgr.  This might be expected following a burst of star formation near
[Fe/H] $-$0.7 to $-$0.6 dex, and a trickle of stars to higher [Fe/H], where
the composition is dominated by AGB stars from the peak of the main burst of star formation.
This would tend to increase the yield of the second,
or heavy, s-process peak elements, like La, qualitatively consistent with the large La
over-abundances toward increasing [Fe/H] and the [La/Eu] versus [La/H] trend seen in 
Figure~\ref{fig-laeulah}.

If the relatively low [La/Y] ratios of MS05/SM02 are taken at face value, then chemical
enrichment from AGB stars with [Fe/H]$>$$-$0.5 dex could explain the [La/Y] ratios 
and the roughly constant [La/Y] value toward higher metallicities.  However, at the present 
time the weight of the published abundances are discordant with the MS05/SM02 [La/Y] values.

We note that abundance measurements of stars in the LMC, by Pompeia et al. (2008)
and Van der Swaelmen et al. (2012), also show large [La/Y] over-abundances, up to $+$1.0 dex,
but with the trend shifted to lower [La/H].  Our dilution calculations indicate that the
Pompeia et al. (2008) trend of [La/Y] with [La/H] in the LMC is consistent
with an AGB dilution curve with [hs/ls]$\sim$$+$1.1 dex, which confirms our conclusion
for a higher than predicted [hs/ls] in Sgr.  

Given the relatively good agreement between the Cristallo et al. (2009, 2011) s-process
predictions and the abundances of MW stars enhanced with AGB-processed material, it
is possible that we have identified a real difference between the [La/Y] ratios in the MW
and the Sgr and LMC dwarf galaxies.  However, we do note the existence of at
least two MW stars with [La/Y] larger than the predictions, for example: the metal-poor CH
stars HE~0024--2523 (Lucatello et al. 2003) and G~24--25 (Liu et al. 2012), both with
[La/Y]=$+$0.85 dex.

While Sgr is enhanced in s-process material that has been ejected at the end of the AGB phase,
we note that comparisons of theoretical s-process predictions with measured heavy element
abundances in MW stars has relied upon current AGB stars and on mass-transfer objects, neither
of which could have reached the final AGB s-process yields.  It seems possible that this
may explain part of the discrepancy in [La/Y] for Sgr and the MW.


We note that the [La/Zr] ratios are also enhanced, and provide the same conclusion that metal-poor
AGB stars contributed significantly to the material of the metal-rich Sgr stars.  However, 
the [La/Zr] for star~247 is somewhat lower than our other two stars, indicating a declining
[La/Zr] with increasing [Fe/H].
On the other hand, the [La/Rb] ratio increases with increasing [Fe/H], similar to the 
[La/Y] trend.  Thus, we have to admit that we do not fully understand the [hs/ls] trends as
well as we would like; however, for Rb, Y, and Zr the [hs/ls] ratios are super-solar,
indicating that the metal-rich Sgr stars formed out of material dominated
by metal-poor AGB stars.


\subsubsection{Rubidium}

Rubidium is thought to be strongly over-produced
by intermediate-mass AGB stars (roughly 4--8M$_{\odot}$) which experience high neutron fluxes
via the $^{22}$Ne($\alpha$,n)$^{25}$Mg reaction.  

The s-process Rb yield is very sensitive to the neutron density, due to an unstable controlling
isotope, $^{85}$Kr, which blocks the production of $^{87}$Rb at low neutron density.
For intermediate-mass AGB stars temperatures are relatively high and $^{22}$Ne($\alpha$,n)$^{25}$Mg
occurs more readily, thus increasing the neutron density.  At these high neutron densities $^{87}$Rb is
produced, but due to its small neutron-capture cross section it is not readily destroyed, resulting in
a large equilibrium Rb abundance.  This is the reason that [Rb/Zr] yields have been assumed to
increase with increasing AGB mass (e.g., see the yields predicted in Smith et al. 2000).


Observationally, large Rb over-abundances
(up to 2 to 5 dex) have been claimed for intermediate mass AGB stars in the Galaxy and the LMC
by Garc\'ia-Hern\'andez et al. (2006) and Garc\'ia-Hern\'andez (2011), respectively.  Recent
theoretical calculations by van Raai et al. (2012) and Karakas et al. (2012) produce Rb
enhancements in intermediate mass AGB stars near $\sim$1 dex, which is similar to the mean
for Galactic Rb-rich stars.  They also find larger Rb-enhancements at lower metallicity,
as observed, but the maximum predicted [Rb/Fe] ratio, thus far, is $+$1.44 dex, significantly
smaller than the most extreme observations.

Observational work and theoretical predictions for lower-mass (1.3--3 M$_{\odot}$) stars 
indicate that the s-process neutrons are provided by the $^{13}$C($\alpha$,n)$^{16}$O 
reaction (e.g., Lambert et al. 1995).  In this case the $^{13}$C 
is produced via $^{12}$C(p,$\gamma$)$^{13}$C following the ingestion of protons from the envelope 
into the He-H intershell region (e.g., Gallino et al. 1998).  Notable, recent calculations of the 
resultant s-process yields for these stars have been performed by Cristallo et al. (2009) and 
Bisterzo et al. (2010).

Recent theoretical s-process yields for low-mass AGB stars (e.g., Cristallo et al. 2009, 2011;
see the the FRUITY\footnote{FRUITY web site http://fruity.oa-teramo.inaf.it/} database) show 
a decline in [Rb/Zr] with increasing mass for AGB stars with masses ranging from 1.3 to
2.0 M$_{\odot}$; however, above a mass of $\sim$2.0 M$_{\odot}$, the [Rb/Zr] yield increases rapidly.
Thus, there is not a linear increase in [Rb/Zr] with AGB star mass, and any attempt to determine
a mean AGB mass or IMF slope from measured [Rb/Zr] ratios is complicated.



\begin{figure}[h]
\centering
\includegraphics[width=8.0cm]{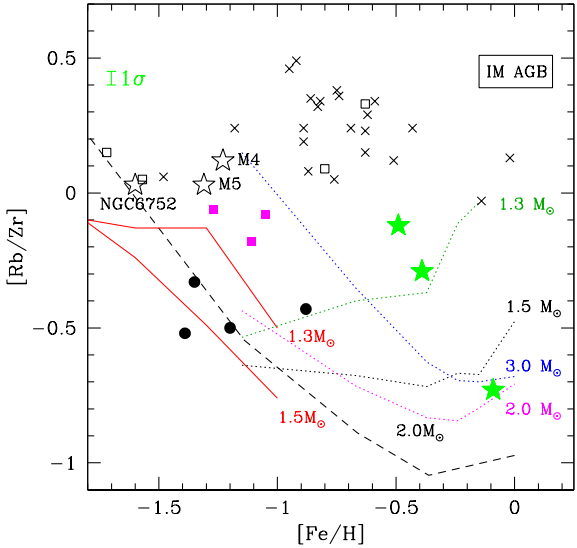}
\caption{[Rb/Zr] for our program stars (filled green stars) compared to the 
Tomkin \& Lambert (1999) solar neighborhood dwarfs and subgiants (black crosses)
and giants (open black squares); CH stars in Tomkin \& Lambert (1999) are indicated
with filled magenta squares.  Galactic globular clusters, M4, M5 and NGC6752, measured
by Yong et al. (2008, 2006) are marked with black open stars. Filled black circles indicate
[Rb/Zr] ratios for $\omega$~Cen, based on Smith et al. (2000), which we have adjusted 
(see text for details).  Also shown are theoretical
predictions for low-mass AGB stars, 1.3 and 1.5 M$_{\odot}$, by Bisterzo et al.
(2010; red solid lines) and 2.0M$_{\odot}$ from Cristallo et al. (2009; black dashed
line).  Dotted lines indicate the latest AGB s-process predictions from the FRUITY 
database.
The theoretical prediction for intermediate-mass (5--7 M$_{\odot}$) AGB stars,
computed by Karakas et al. (2012), is marked with a boxed ``IM AGB''.
\vskip 0.2cm
}
\label{fig-rbzr}
\end{figure}

Figure~\ref{fig-rbzr} shows a comparison of [Rb/Zr] for
our three Sgr stars with the Galactic disk
and halo.  We have scaled the Smith et al. (2000)
$\omega$~Cen [Rb/Zr] points to the solar meteoritic Rb/Zr ratio indicated by Lodders, 
Palme \& Gail (2009), at $-$0.19 dex.
We have also adjusted the [Rb/Zr] ratios downward by 0.09 dex for the
Tomkin \& Lambert (1999) RGB stars, in order to account for their over-estimated EWs of the solar 
Zr~I lines at 6127.5 and 6143.2\AA .  Our inspection of the profiles of these lines
in the Kurucz solar spectrum reveals obvious blends, so our EW measurements are somewhat smaller
than found by Tomkin \& Lambert (1999);  however, we do not see these blends
in the Arcturus atlas of Hinkle et al. (2000).
Although we should probably apply the same downward correction to the [Rb/Zr] ratios
for both the dwarf and turnoff stars in Tomkin \& Lambert (1999), we do not know whether
their normalization relative to the sun causes the effects of the blends to cancel-out in the dwarfs;
therefore, we have not applied corrected the Tomkin \& Lambert (1999) Zr abundances for dwarf stars.

It is clear that the [Rb/Zr] ratios for our Sgr stars are well below 
the solar neighborhood, mostly thick disk, stars of Tomkin \& Lambert (1999);
our [Rb/Zr] ratios are also lower than the globular
clusters, except for $\omega$~Cen, which has similar values.  
The Sgr [Rb/Zr] ratios decline roughly linearly with increasing [Fe/H], indicating the 
inclusion of progressively more material from low-mass AGB stars as metallicity increases.  
Thus, s-process nucleosynthesis in Sgr is driven by the $^{13}$C($\alpha$,n)$^{16}$O
neutron source that operates in low-mass AGB stars.

The linear decline in [Rb/Zr] with increasing [Fe/H] suggests an initial 
[Rb/Zr]$\sim$0.0 dex ratio, similar to the Galactic globular clusters, but near
[Fe/H]=$-$0.6.  This was followed by the addition of progressively more
material from low-mass ($\sim$2 M$_{\odot}$) AGB stars as time and [Fe/H] increased.
Certainly, there was no significant contribution to the [Rb/Zr] ratio in Sgr,
above [Fe/H]=$-$0.5, by intermediate-mass AGB stars.  
If low-mass AGB material dominates in this way then the most s-process enhanced Sgr stars
should show very low, or absent,$^{25}$Mg and$^{26}$Mg isotopes, as evidenced from their
MgH line profiles.

Clearly, more points are required to verify the [Rb/Zr] versus [Fe/H] trend seen here.  The similarity
of the [Rb/Zr] values measured here for Sgr to the values in $\omega$~Cen, found by 
Smith et al. (2000), is yet another chemical signature shared between these two systems
(e.g., McWilliam \& Smecker-Hane 2005a,b).

Lines in Figure~\ref{fig-rbzr} indicate predicted [Rb/Zr] yields for AGB stars as a function of 
metallicity and mass.  The predictions show that intermediate-mass AGB stars (5--7M$_{\odot}$) produce
[Rb/Zr]$\sim$$+$0.4--0.5 dex (e.g. Karakas et al. 2012), while low-mass AGB stars yield much lower
[Rb/Zr] ratios, as low as $-$1.0 dex.  Also, there is a strong increase in [Rb/Zr] with decreasing
[Fe/H] below metallicities corresponding to roughly [Fe/H]=$-$0.5 dex (e..g., Bisterzo et al. 2010;
Cristallo et al. 2009).

These theoretical predictions indicate that the [Rb/Zr] ratio for our most metal-rich Sgr star,
\#~247, is inconsistent with AGB s-process below [Fe/H]$\sim$$-$1.0.
Since the metallicity of M54 is below this metallicity, at [Fe/H]$\sim$$-$1.6 dex
(Brown \& Wallerstein 1999; Carretta et al. 2010), it follows
that the neutron-capture elements in star 247 cannot have been produced 
by a low-mass M54 AGB star.  If dilution is responsible for the observed linear trend of 
[Rb/Zr] with [Fe/H] in our three Sgr stars, then M54 AGB stars could not have been 
responsible for their neutron-capture elements. On the other hand, it is
possible that the [Rb/Zr] ratio in star 247 could have been produced by an AGB star near
[Fe/H]$\sim$$-$0.6 dex, roughly the mean metallicity of Sgr.


The low [Rb/Zr] ratios in Figure~\ref{fig-rbzr} indicate that the ratio of intermediate-mass
to low-mass AGB stars in Sgr must have been much lower than for the MW disk and halo.
This enhanced enrichment from low-mass AGB stars is consistent with the large s-process
enhancements already noted.  It is possible that these relatively low [Rb/Zr] ratios could 
result from an IMF that is heavily weighted to the lowest mass stars (a bottom-heavy IMF);
however, such a conclusion regarding the IMF is not yet warranted. 

Following a burst of star formation, material from low-mass AGB stars can dominate the
gas composition at later times if a significant amount of the gas is lost from the system
after the burst, i.e., leaky-box chemical evolution.   Such outflows were suggested by SM02
to explain the high [La/Y] ratios (characteristic of low metallicity AGB stars) found even 
for the highest metallicity Sgr stars.  An important point is that the [La/Y] ratios require
the dominance of metal-poor AGB material from the burst of star formation (near [Fe/H]$\sim$$-$0.6
dex) to the late-time, higher-metallicity, gas.  The Sgr metallicity distribution function
(e.g. Bellazzini et al. 2008; C10) and the Sgr mean [Fe/H] being lower than the MW,
is qualitatively consistent with a loss of gas reducing the formation of higher metallicity
stars.

Even a system that does not leak, but does not experience significant gas inflows
could be expected to show lower [Rb/Zr] ratios than the MW disk, which is characterized
by continuous star formation and gas infall.

In one version of a leaky box scenario, following a burst of star formation, at any
time after the burst the chemical composition of the gas is dominated by the stars from the
burst that are currently ejecting their envelopes.  Thus, as time and metallicity increase
the gas composition, and the composition of the diminishing younger stellar populations,
is dominated by the ejecta of progressively lower mass, older, stars.  The exact mix
of older and younger material would depend, in part, on the rate of gas leakage
from the galaxy and the amount of star formation following the main event.

Detailed chemical evolution models fit to the measured chemical abundance 
patterns will be required to determine whether a bottom-heavy IMF or leaky-box 
model better explains the low [Rb/Zr] ratios in Sgr; however, we favor the leaky box scenario
because outflows are more likely in low-mass galaxies, like Sgr, whose gravity is less able
to retain hot or high-velocity supernova ejecta.

Regarding timescales: from Figure~\ref{fig-rbzr} it is clear that the low [Rb/Zr] ratio for
star 247 could have been produced by a 2 M$_{\odot}$ AGB star; the main sequence lifetime
of such stars is $\sim$1 Gyr (e.g., Pietrinferni et al. 2004).  Thus, there was plenty of time
between the 4--6 Gyr and 2.3 Gyr star formation bursts in Sgr (Siegel et al. 2007),
to reduce the [Rb/Zr] ratio with material from low-mass AGB stars.  The 2M$_{\odot}$ AGB stars
might even have enriched Sgr with s-process elements {\em during} the burst
of star formation 4--6 Gyr ago.
Notably, the age gap between the 4--6 Gyr and 2.3 Gyr populations indicates sufficient time to permit
the lowest-mass s-process 
producing AGB stars, at 1.3 M$_{\odot}$ (e.g., Busso et al. 2004), with main-sequence lifetimes of 
$\sim$3 Gyr, to enrich the late-time gas in Sgr.

\subsection{The r-Process and [Eu/O]}

\begin{figure}[h]
\centering
\includegraphics[width=8.0cm]{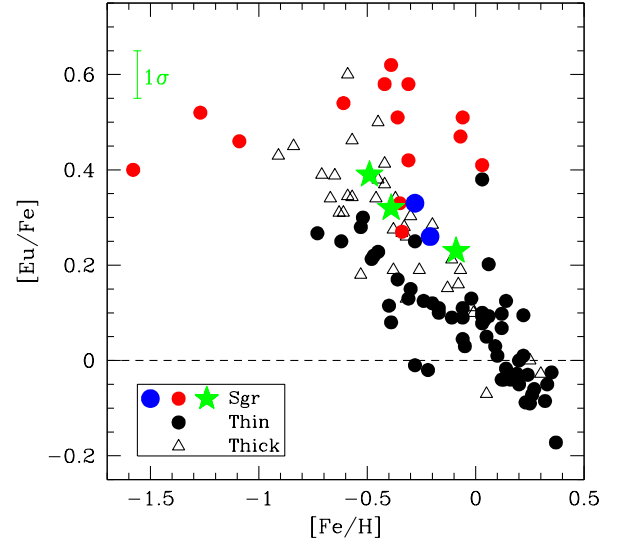}
\caption{
A plot showing the [Eu/Fe] ratio versus [Fe/H] for stars in the
Galactic thin and thick disk (black filled circles and open triangles respectively)
from Bensby et al. (2005) compared with measurements of Sgr stars: filled green stars 
(this work), filled red circles  (SM02) and blue open circle (B00).
}
\label{fig-eufe_obs}
\end{figure}

In Figure~\ref{fig-eufe_obs} we show the observed [Eu/Fe] versus [Fe/H] for the three
stars studied here, compared to results from B00 and 
SM02; the SM02 Eu abundances have been reduced downward by 0.08 dex to correct an
arithmetic error in the original work.  We also compare to the [Eu/Fe] ratios for
the Galactic thin and thick disks from Bensby et al. (2005).

McWilliam \& Smecker-Hane (2005a) found enhanced [Eu/Fe] ratios in Sgr, but
concluded that these could reasonably be due to s-process contributions to the Eu abundances.
While their argument was plausible, based on the very strong La and Ba lines that indicated
strong s-process enhancements, they did not compute the Eu s- and r-process fractions in
their Sgr stars.  Here, we compute the Eu r-process fractions, f(Eu)$_r$, for our three 
stars, using Equation~3.


\begin{equation}
f(Eu)_r = \frac{1 - R_* / R_s} {1 - R_r/R_s}
\end{equation}

Where
$$ R_* = \bigg(\frac{N(La)}{N(Eu)}\bigg)_*  =  10^{\varepsilon(La)_* - \varepsilon(Eu)_*}  $$
$$ R_r = \bigg(\frac{N(La)}{N(Eu)}\bigg)_r  =  10^{\varepsilon(La)_r - \varepsilon(Eu)_r}  $$
$$ R_s = \bigg(\frac{N(La)}{N(Eu)}\bigg)_s  =  10^{\varepsilon(La)_s - \varepsilon(Eu)_s}  $$

Subscripts ``s'' and ``r'' refer to pure s-process and r-process values, respectively; the
asterisk indicates values for the star under consideration.

Critial inputs to Equation~3 are the pure r- and s-process La/Eu number ratios.  The r-process
[La/Eu] ratio is taken from residuals to the solar system N$\sigma$ curve by K\"appeler et al. (1989),
Burris et al. (2000), and Simmerer et al. (2004), giving [La/Eu]$_r$ of $-$0.58 dex, on the Lodders et al.
(2009) meteoritic scale (where $\epsilon$(La/Eu)$_{\odot}$=$+$0.66 dex);
this is identical to the [La/Eu] abundance ratio, measured by Sneden et al. (1996), for the
r-process rich star CS~22892-052.  

The s-process ratio is more difficult to assign, mainly because the exact [La/Eu] value
depends upon the details of the s-process site, including the stellar mass and metallicity.  For example,
a strong s-process computed by Malaney (1987) gave [La/Eu]$_s$=$+$0.85 dex for a single neutron exposure 
of $\tau$ = 1.5 mb$^{-1}$; much more recently,
Cristallo et al. (2009) predicted [La/Eu]$_s$=$+$0.877, $+$1.076, and $+$0.95 dex for 2M$_{\odot}$
AGB stars with metallicities of 0.0, $-$0.36 and $-$0.66 dex, respectively; other AGB predictions,
by Arlandini et al. (1999) and Bisterzo et al. (2010) give similar s-process [La/Eu] values.  Thus,
it appears that there is general agreement on the theoretically predicted s-process ratios, 
near +0.88 dex; however, the N$\sigma$ fits to the solar abundance distribution, by 
Burris et al. (2000) and Simmerer et al. (2004) gave significantly higher ratios, both indicating
[La/Eu]$_s$=$+$1.47 dex.

Our computed Eu r-process fractions, f(Eu)$_r$, based on the lowest theoretical [La/Eu]$_s$ values,
near $+$0.88 dex (Cristallo et al. 2009), are 0.86, 0.79 and 0.87 for stars 242, 247 and 266 respectively.
If we adopt [La/Eu]$_s$=$+$1.47 from Simmerer et al. (2004) and Burris et al. (2000) we obtain 
f(Eu)$_r$ values of 0.97, 0.95, and 0.97 respectively.  Thus, it is clear that despite the s-process
enhancements in our Sgr stars, their Eu abundances are still dominated by the r-process.  In the worst
case the maximum correction to apply to the total Eu abundance, in order to obtain the r-process Eu
abundance, is $-$0.10 dex.  

\begin{figure}[h]
\centering
\includegraphics[width=8.0cm]{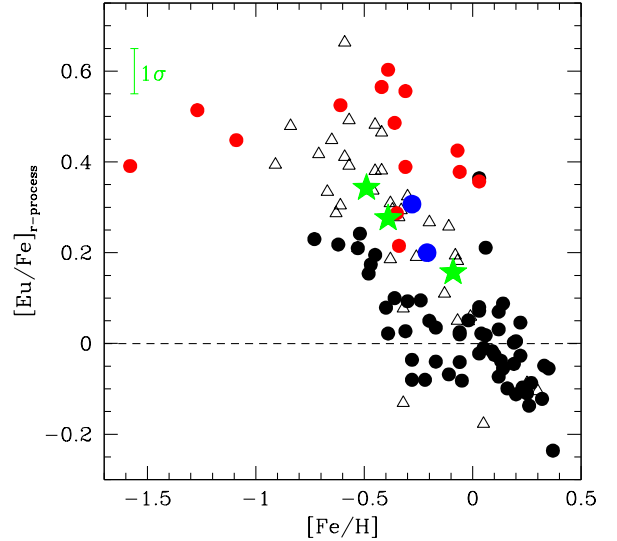}
\caption{
A plot showing the r-process [Eu/Fe]$_r$ ratio versus [Fe/H] for stars in the
Galactic thin and thick disk (black filled circles and open triangles respectively)
from Bensby et al. (2005) compared with Sgr: filled green stars (this work),
filled red circles (SM02), and blue open circle (B00).  See the text for a description
on how the s-process fraction was subtracted in order to obtain the r-process, [Eu/Fe]$_r$,
values.
}
\label{fig-eufer}
\end{figure}

In the following discussion we adopt a compromise value of [La/Eu]$_s$=$+$1.00 dex;
for this case we find f(Eu)$_r$ of 0.90, 0.84 and 0.90 for stars 242,
247, and 266 respectively.  We apply Equation~3 with the same [La/Eu]$_s$ to compute f(Eu)$_r$
for Eu abundances in SM02 and B00; we also add the $-$0.08 dex correction
to Eu abundances in SM02 to correct an arithmetic error in their work.  To compute
r-process Eu abundances we simply add log$_{10}$ f(Eu)$_r$ to the measured Eu abundances.  The largest
correction to SM02 Eu abundances is for their most metal-rich, and La-rich star, at [La/Fe]=$+$0.96,
with a $\varepsilon$(Eu) correction of $-$0.13 dex.  Typical r-process corrections for SM02 Eu abundances were
a mere $-$0.02 dex.  When these corrections were applied to this work and to B00 for Sgr, and
to the Bensby et al. (2005) MW disk results, our [Eu/Fe]$_r$ values, and those of B00, lie
in excellent agreement with the MW thick disk values, as seen in Figure~\ref{fig-eufer}; however,
the SM02 results still lie above the MW [Eu/Fe]$_r$ trend with [Fe/H].

The similarity between the MW disk and Sgr r-process [Eu/Fe]$_r$ trend with [Fe/H] is
particularly striking in comparison to the deficient trend of [O/Fe] with [Fe/H] in Sgr.
It is not possible to explain both the decline of [O/Fe] and [Eu/Fe]$_r$ in Sgr with the late 
addition of iron from Type~Ia supernovae, as used by Tinsley (1979) to explain the [O/Fe] trend in
the MW disk.

\begin{figure}[h]
\centering
\includegraphics[width=8.0cm]{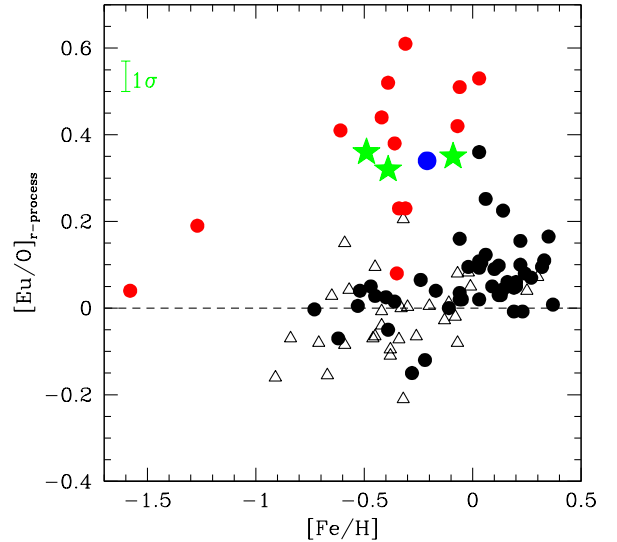}
\caption{
A plot showing the r-process [Eu/O]$_r$ ratio versus [Fe/H] for stars in the
Galactic thin and thick disk (black filled circles and open triangles respectively)
from Bensby et al. (2005) compared with Sgr stars: filled green stars (this work),
filled red circles (SM02/MS05a), and blue open circle (B00).
The Sgr stars shows an [Eu/O]$_r$ enhancement of $\sim$0.35--0.40 dex,
while the disk ratio is relatively constant over 1.5 dex in [Fe/H].
Note that pure s-process [La/Eu]=$+$1.0 was assumed in the calculation of the europium
r-process fraction, f(Eu)$_r$.
}
\label{fig-euor}
\end{figure}

\begin{figure}[h]
\centering
\includegraphics[width=8.0cm]{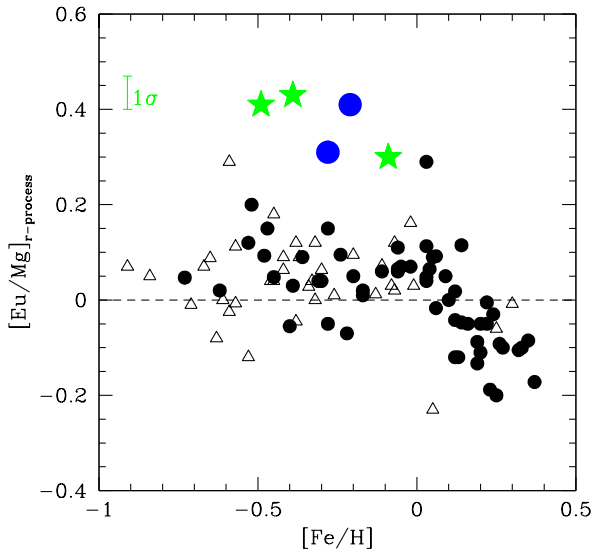}
\caption{
A plot showing the r-process [Eu/Mg]$_r$ ratio versus [Fe/H] for stars in the
Galactic thin and thick disk (black filled circles and open triangles respectively)
from Bensby et al. (2005) compared with Sgr: filled green stars (this work)
and blue open circles (B00).  Sgr shows an [Eu/Mg]$_r$ enhancement
of $\sim$0.40 dex.
Note that pure s-process [La/Eu]=$+$1.0 was assumed in the calculation of the europium
r-process fraction, f(Eu)$_r$.  
}
\label{fig-eumgr}
\end{figure}

In Figures~\ref{fig-euor} and \ref{fig-eumgr} we show the r-process [Eu/O]$_r$ and [Eu/Mg]$_r$ for Sgr 
stars in this work, SM02 and B00, compared with the MW thin and thick
disk results of Bensby et al. (2005).  Mg and O are thought to be produced mainly in the hydrostatic
burning phases of massive stars that end as SNII.
The filled black circles and open black triangles in Figure~\ref{fig-euor}
indicate the thin and thick disk stars, respectively.  As noted by Bensby et
al. (2005), the MW [Eu/O] ratio is flat, near the solar value, over more than 1 dex in [Fe/H], 
suggesting that Eu and O were formed in similar environments.  This is consistent with 
the widely accepted idea that the r-process likely occurs in SNII; certainly, the r-process
timescale, of $\sim$1 second, suggests a sudden dramatic event, typical of supernovae.  However, 
theoretical investigations into the r-process have yet to identify a viable mechanism
(e.g., see Nishimura et al. 2012).  Figures~\ref{fig-euor} and \ref{fig-eumgr} show that the 
r-process [Eu/O]$_r$
and [Eu/Mg]$_r$ ratios in the bulk of Sgr stars are enhanced by $\sim$0.35 dex relative to
the Galactic thin and thick disks.  This is confirmed by the SM02/MS05a and Bonifacio et al. (2000)
Sgr abundances.

How could the r-process [Eu/O]$_r$ ratios be so consistent in the Galactic disk but enhanced in the Sgr?
We suggest that these [Eu/O]$_r$ enhancements result from a paucity of the most massive SNII
progenitor stars (which dominate oxygen production), and that the r-process is produced mostly in SNII
with masses lower than the average 
O and Mg producers.  This paucity of the most massive high-mass stars could be due either to a steep
massive-star IMF, or a ``top-light'' SN mass function missing the upper mass range.

Abundance studies of extreme metal-poor MW halo stars indicate that there exists a large
range of r-process yields from SNII and that the bulk of the r-process elements were produced
in a rare SNII sub-type, at most only a few percent of all SNII (e.g., McWilliam et al. 1995;
Sneden et al. 1994, 1996; McWilliam 1998; McWilliam \& Searle 1999; Fields et al. 2002).  

We do not yet know the controlling factor, or factors, responsible for this rare, r-process, 
SNII sub-type.
However, the distinctly different [Eu/O]$_r$ ratios in Sgr and MW disks
shows that there is a systematic difference in one, or more, characteristics of the SNII
between these two systems.  While angular momentum of the SNII progenitor and binary
membership might play a role in r-process production, we think it unlikely that these
parameters differ significantly between the MW disks and Sgr.  Similarly, metallicity 
is probably not the parameter mainly responsible for our [Eu/O]$_r$ result, because our Sgr 
stars and the MW thick disk sample share a similar metallicity range.  However, it seems
possible that metallicity could play a role in modulating r-process yields, in general.  On the
other hand, the mass function of SNII
progenitors could well play a role in total r-process production and could reasonably differ
between the MW disk and Sgr.

Under the assumption that the r-process yields are sensitive to the SNII progenitor masses,
the enhanced [Eu/O]$_r$ ratio indicates that the mass function of SNII progenitors
is more strongly weighted toward the mass of the r-processing producing SNII in Sgr than
in the MW disk.  For example, if r-process producing SNII are predominantly lower-mass then
[Eu/O]$_r$ enhancement would occur if the IMF is weighted to lower-mass SNII progenitors  (i.e., a steep 
upper-end IMF slope or an upper mass cut-off).  

The enhanced [Eu/O]$_r$ ratios, alone, cannot determine whether the IMF is steeper or shallower
than Salpeter; although, the fact that oxygen is produced preferentially in high-mass SNII
suggests a steeper IMF in Sgr, i.e., with a paucity of high-mass stars.  Additional evidence 
for such a ``top-light'' IMF in Sgr include the deficiencies of hydrostatic elements, O, Na, Mg, Al and Cu
relative to explosive elements Si, Ca and Ti, consistent with muted nucleosynthesis contribution from 
SNII compared to SNIa.
The integrated-galactic IMF (IGIMF) has been predicted to be steeper in dwarf galaxies, like
Sgr (e.g., Weidner \& Kroupa 2005; Kroupa et al. 2011; see also Oey 2011).  The reason is that dwarf
galaxies have, by definition, less gas mass and lower-mass molecular clouds than normal size galaxies, 
and as a result they are less efficient at producing the highest mass stars.  Kroupa et al. (2011) also
predicted a metallicity-dependence of the IGIMF slope, which they expected to steepen with
increasing metallicity.

Thus, the measured enhanced [Eu/O]$_r$ ratios leads two interesting conclusions: 1) That r-process
SNII are lower mass than the typical
oxygen-producing and magnesium-producing SNII, and 2) That either the massive-star IMF in Sgr
was steeper than the MW disk, or there was an cut-off in the upper-end IMF, such as might
be expected from Weidner \& Kroupa (2005) and Oey (2011) IGIMF expectations.




If these conclusions are correct, then other dwarf galaxies should also show enhanced [Eu/O]$_r$ and [Eu/Mg]$_r$
ratios, since they, too, should possess IGIMF steeper than the Salpeter value.  
Stars in Fornax (Letarte et al. 2010, 2012), the LMC (Mucciarelli et al. 2008, 2010; Andrievsky et al. 2001;
see also Van der Swaelmen et al. 2012),
and IC~1613 (Tautva$\breve{\rm s}$ien$\dot{\rm e}$ et al. 2007) show similar enhancements in [Eu/O]$_r$ 
and [Eu/Mg]$_r$ to 
those in Sgr.  However, the relatively metal-poor dwarf galaxies Sculptor (Shetrone et al. 2003;
Geisler et al. 2005) and Carina (Shetrone et al. 2003; Venn et al. 2012) do not share these enhancements.
This non-uniform behavior might be explained by the predicted metallicity-dependence of the IGIMF.  
Systematic measurement error in the Eu abundances of metal-rich stars seems unlikely, as the same
problem would have occurred in MW disk red giants, but this is not seen.

\subsection{Alternative Scenarios?}

We have found that the declining [O/Fe] and [Eu/Fe]$_r$ trends with [Fe/H] in Sgr cannot both be explained
by the late addition of Fe from delayed SNIa; two mechanisms are necessary.  We have proposed a top-light
or steep IMF in Sgr to reduce the abundances of the hydrostatic (O, Mg, Na, Al, Cu) and explosive (Si, Ca, Ti)
element families; the [Eu/Fe]$r$ trend could then be explained by the assumed delayed addition of Fe from SNIa,
following the Tinsley (1979) scenario.  If this is correct, then Sgr and the MW thick disk must have similar
SFR, because the [Eu/Fe]$_r$ trend is similar in both systems; this may not be an unreasonable assumption,
based on approximate age ranges of thick disk and Sgr populations.

An alternative idea is that the hydrostatic and explosive element family is deficient, relative to Fe,
due to a low SFR and extra Fe from SNIa in Sgr, following Tinsley (1979) and Matteucci \& Brocato (1990),
with the [Eu/Fe]$_r$ decline due to metallicity-dependent r-process yields.

The problem with this idea is 
that the Sgr and MW [Eu/Fe]$_r$ trends are very similar, yet the time-delay mechanism suggests more SNIa
Fe in Sgr than in the MW.  To accomplish this requires that the extra SNIa Fe in Sgr is accompanied by
extra Eu production, exactly enough to make the Sgr and MW [Eu/Fe]$_r$ trends with [Fe/H] look similar.  
The r-process would be made by both SNIa and SNII in this model, which may be difficult to understand
the light of the large dispersion of [Eu/Fe]$_r$ in the metal-poor halo (e.g., McWilliam et al. 1995),
indicating that the r-process is made in only a few percent of all supernovae.  
However, this mechanism faces the fatal question of how the Sgr SNIa at [Fe/H]=$-$0.5 dex, say,
know to make extra Eu, but the MW SNIa at the same metallicity do not make this extra Eu.  This seems
impossible.  

Thus, this particular alternate scenario: where the Sgr hydrostatic X/Fe ratios were reduced by delayed SNIa Fe
production at low SFR with Eu/Fe declining due to metal-dependent yields, and similar to the MW disk,
introduces more problems than it solves.

In the context of the time-delay scenario, it seems much more reasonable to explain our measured
abundance ratios by appealing to an IMF deficient in the most massive SNII progenitors with the
r-process located in lower-mass SNII, for which other evidence and arguments already exist.

We emphasize that metal-dependent r-process yields can be made consistent with our Sgr measurements and 
the MW if the time-delay mechanism is abandoned.  In such a model hydrostatic, explosive and r-process
trends with [Fe/H] would all largely result from metallicity-dependent effects.  This possibility will
be discussed in more detail in a forthcoming paper (McWilliam 2013).

%

\section{The Chemical Evolution of Sgr}

Various element abundance ratios provide clues to the chemical
evolution of Sgr.  The most simple diagnostic is [Fe/H], often
incorrectly referred to as the ``metallicity''.  In Sgr the mean
[Fe/H] is near $-$0.5 to $-$0.7 dex (Cole 2001; Bellazzini et al. 2008; 
Siegel et al. 2007), much lower than the thin disk of the MW.  

The low average [Fe/H] in Sgr may be due to significant mass-loss during its
evolution, which truncated chemical enrichment before complete conversion of 
the gas into stars could occur.  
A similar gas-loss mechanism was proposed to explain 
the low average [Fe/H] of the MW halo by Hartwick (1976).  Also, Kirby et al. (2013)
required significant outflows to understand the mean metallicities of a sample 
of Local Group dwarf galaxies.

Presumably, outflows are more likely in low-mass galaxies, like Sgr, than the MW, because
the relatively shallow gravitational potential well would permit more high-velocity
supernova ejecta to leave than for massive galaxies.  However, tidal stripping by
interaction with the MW must also have been an important gas-loss mechanism for Sgr, as
evidenced by the prominent stellar tidal tails.  Sgr contains almost no gas today.  

A variant on overall mass-loss is selective mass-loss, for example energetic
material from SNIa or SNII ejecta are, presumably, more likely to be ejected from 
Sgr than  the low-velocity ejecta from planetary nebulae (PNe).  Thus, one might reasonably
expect a larger chemical abundance signal from PNe in dwarf galaxies; indeed, this may
be consistent with the enhanced s-process abundances in Sgr and other dwarf galaxies.
In terms of overall metallicity, selective mass-loss from SNII and/or SNIa would
lower the yield of metals per stellar generation and reduce the amount of gas to
be recycled into stars; both mechanisms would result in a lower mean [Fe/H].

A low mean [Fe/H] can also result from a steep stellar IMF.
In a closed box model of chemical evolution, the average metallicity of 
the system after all the gas is consumed is equal to the yield (Searle \& Sargent 1972),
a ratio which is the mass of metals produced divided by the
mass locked-up in low mass stars.  A system with a steep IMF slope
(i.e. with a relatively high frequency of low-mass dwarfs) will, therefore, have
a smaller yield, and the mean metallicity will be lower than for a
normal IMF slope.  In other words: for a steep IMF slope more gas is locked-up
in low mass stars, leaving fewer high-mass stars to produce the metals and thus,
the final average metallicity is lower.

Why did the Sgr metallicity distribution function (henceforth MDF) turn-over near [Fe/H]=$-$0.6 dex?  
Was it due to a loss
of gas from the shallow gravitational potential of Sgr, or some self-limitation
on the star formation rate, possibly due to gas heating from hot stars; or did 
the MDF turnover result from a low effective yield, as a consequence of
a steep IMF slope, or the selective loss of metals, or from a
low binary fraction, leading to fewer SNIa events?

In light of Sgr's low mass, perhaps the most natural answer to this question is that its low mean
[Fe/H] is due to chemical evolution in the presence of significant gas loss, i.e., leaky box
evolution, as suggested by SM02 and MS05.

Our IMF diagnostics include the hydrostatic and explosively produced elements,
O, Mg, Na, Al, Cu, Si, Ca, Ti and Eu.  As outlined earlier in this paper, 
the ratio of hydrostatic to explosive element abundances,
and in particular the [Eu/O] and [Eu/Mg] ratios, indicate a deficit of high-mass SNII, most 
likely due to a top-light integrated galactic IMF (IGIMF).  This probably resulted from the limited
amount of gas available in Sgr, that restricted the formation of the most massive molecular clouds,
from which the most massive stars are born (e.g., Weidner \& Kroupa 2005; Kroupa \& Weidner 2003).

Given the similarity of the chemical abundance ratios in Sgr with the LMC, it is reasonable 
to suppose a similar origin for the measured abundance ratios, in which case the LMC should
also suffer from a top-light IGIMF.  While the IMF slopes for clusters in the LMC are
close to the Salpeter value seen in the MW (e.g. Massey 2003), the 
IMF derived from the measured current mass function of LMC field stars,
a measure of the IGIMF, shows steep slopes, $\Gamma$$\sim$$-$4, above 1M$_{\odot}$ 
(Gouliermis, Brandner, \& Henning 2006; Massey et al. 1995), consistent 
with a top-light IGIMF.  
However, measured mass functions for field stars more massive than $\sim$1M$_{\odot}$
require corrections to account for evolved stars and past star formation, in order to 
estimate the IMF.  

We note that IMF slopes for MW field
star slopes are also steep for masses 1$<$M$<$4 M$_{\odot}$, similar to the LMC values
(e.g., Reid, Gizis \& Hawley 2002); although, we know of no MW measurements for the field-star IMF
above 4M$_{\odot}$.
At present it is not possible to say whether the measured IMF slope for massive stars in the LMC
field is steeper than the MW; see reviews by Massey (2003), Elemgreen (2004), Kroupa et al. (2011), 
and Scalo (2005) for further discussion.  For the SMC field star IMF Lamb et al. (2013) found a
slope significantly steeper than Salpeter, at $\Gamma$=$-$2.3$\pm$0.4, for masses below
$\sim$7 M$_{\odot}$; although, the lowest two mass bins the slope were consistent with the Salpeter IMF slope.
Numerical experiments by Lamb et al. (2013) showed that star formation corrections would most likely
not affect the measured field-star IMF slopes.  Thus, for the SMC a steep IMF is also indicated.  

A comparison of the OB associations
in M31 and the Local Group dwarf galaxy NGC6822, by Massey et al. (1995b) showed that NGC6822
OB associations have significantly fewer OB stars, with luminosities much less than in M31.
This provides a crude comparison of the IGIMFs for the two galaxies, and is consistent with, but
does not prove, the existence of a top-light IGIMF in NGC6822.

We conclude that there is mounting evidence that Local Group dwarf galaxies actually do possess steep
IGIMF slopes, qualitatively consistent with the chemical abundance ratios measured in this work.
However, the interpretation is complicated by necessary corrections and the fact that field stars
in the MW disk also possess a steep present-day IMF slope.


As mentioned earlier, steep IGIMF slopes have been predicted for dwarf galaxies
(e.g., Weidner \& Kroupa 2005) where there is insufficient gas to form the most massive
molecular clouds, and hence the most massive stars.  If this is true, then one should
expect steep IMFs to occur whenever a system runs low on gas, such as in the presence of strong
outflows, or simply gas consumption due to star formation.  In the latter case, there will be
relatively
little star formation, and the IMF steepening will occur after the peak of the MDF.  Indeed, it 
might be generally expected that the IMF slope will be steeper in the high-metallicity tail of
the MDF.  In this regard, it is interesting that the MW disk [Mg/Ca] ratios decline with increasing 
[Fe/H], although this might be due to increasing Ca from SNIa.  


The large s-process enhancements in Sgr, such as [La/Fe], are evidence
of the importance of low-mass AGB stars in the chemical evolution of this galaxy. 
This might be used to argue for a steep IGIMF slope, to low masses, in Sgr.
However, the
[La/Y] ratios in Sgr show, very clearly, that low-metallicity AGB stars produced the neutron-capture
abundance patterns seen at high metallicity.
Unsurprisingly, there was not instantaneous recycling of material in Sgr.  The [La/Y] ratios
show that the evolved low-metallicity stars from the peak of the [Fe/H] distribution function
dominated the later evolution.  Here we follow the suggestion of SM02 and MS05: that outflows,
or a leaky box scenario, caused the [La/Y] ratios of the metal-poor Sgr stars to dominate
at high [Fe/H].  At any time, the gas was dominated by the ejecta from the peak of the
metallicity function, near [Fe/H]=$-$0.6 dex, but due to progressively lower mass stars.

Our very low [Rb/Zr] ratios suggest that 2M$_{\odot}$ AGB stars, with [Fe/H]$\sim$$-$0.6, dominated
the chemical composition of the Sgr gas at [Fe/H]$\sim$$-$0.1 dex.  Because the AGB s-process
nucleosynthesis predictions of Cristallo et al. (2009, 2011) and Bisterzo et al. (2010) 
produce the very low [Rb/Zr] ratios seen in Sgr only for [Fe/H] above $-$1 dex,
the source of the AGB material cannot have been from a population with lower
metallicity, such as a metal-poor halo or M54, which has [Fe/H]$\sim$$-$1.8 dex.
Thus, the region of Sgr studied here was likely self-enriched without
significant inflow of metal-poor gas from other parts of the galaxy.

In summary, the abundance ratios measured here, and in other studies of Sgr, indicate
chemical evolution of a system with a top-light IMF in the presence of outflows.
The outflows are mainly responsible for the low mean metallicity of Sgr,
and cause a steady reduction in the population with increasing time and metallicity.
Thus, ejecta from the relatively small metal-rich population is overwhelmed by nucleosynthesis
products of the old [Fe/H]$\sim$$-$0.6 AGB stars.  The top-light IMF
results in abundance deficiencies of hydrostatic elements, like O, Mg, Na, Al,
and Cu, but smaller deficiencies of explosive alphas, like Si, Ca, and Ti.  
The [Eu/Fe] trend is similar to the MW disk; if we assume the time-delay paradigm of
Tinsley (1979), this suggests that the SFR in Sgr was similar to the MW disk.

A test of this leaky-box scenario, in which the metal-rich composition was dominated by the products
of metal-poor stars, may be facilitated by the trend of [Mn/Fe] with [Fe/H].  If metal-poor SNIa, following
a significant time delay, contribute iron-peak elements to the metal-rich Sgr gas, then deficient [Mn/Fe]
ratios are expected.  Presently, there is dispersion in the reported [Mn/Fe] ratios from different
Sgr abundance studies, so further investigation is required.

\section{Summary and Conclusions}

  We have performed a high-resolution abundance analysis of 3 stars on the faint RGB
  towards M54.  This giant branch has previously been shown to belong to 
  Sgr, kinematically and chemically distinct from M54.

  Our measurements indicate [Fe/H] values of $-$0.49, $-$0.39 and $-$0.09 dex for our
  three stars, consistent with previous measurement of the [Fe/H] distribution 
  in Sgr.  Our velocity dispersion is more consistent with the
  Sgr value indicated by Bellazzini et al. (2008), and less consistent M54 kinematics;
  however, with only 3 points we can make no strong conclusion regarding the velocity dispersion
  of our stars.

  Our detailed chemical abundance ratios are consistent with previous studies of
  stars in Sgr: deficient [$\alpha$/Fe] ratios, particularly [O/Fe] and [Mg/Fe], 
  suggest relatively less enrichment by the most massive SNII; significant deficiencies of
  [Na/Fe], [Al/Fe], and [Cu/Fe] also indicate a paucity of ejecta from more massive SNII;
  enhancements of neutron-capture elements, made by the s-process, show enrichment by AGB
  stars, with the enhanced heavy to light s-process ratio (e.g., [La/Y]) consistent with
  low-metallicity AGB nucleosynthesis.  These unusual abundance ratios provide further evidence 
  that our program stars are, indeed, members of Sgr.  Furthermore, our results and those of S07
  show no evidence of an Na--O abundance anti-correlation, seen in most globular clusters, which
  supports the conclusion that our stars are not members of M54.

 We do not confirm the deficient [V/Fe] ratios found by Sbordone et al. (2007), nor the
 Na enhancements found for the metal-rich Sgr stars by Carretta et al. (2010).  We
 believe these earlier claims to be spurious, arising from systematic errors in these other
 studies, perhaps due to model atmosphere problems and blends. However, given the 
 spatial distance between the S07 sample and those of C10 and SM02 it is possible that
 a chemical abundance gradient in could have produced the putative vanadium deficiency claimed
 by S07.  

We also do not confirm the $\sim$0.2 dex 
 lower-than-normal [Mn/Fe] values claimed by McWilliam et al. (2003); in this work we find 
 the [Mn/Fe] trend resembles the MW disk and halo.  
This difference may be due to the uncertainties involved with the
 absolute abundance technique of SM02 and McWilliam et al. (2003) combined with the paucity of
 photometric and reddening information available, which made it difficult to
 constrain the atmosphere parameters of their stars.  
 Additionally, in SM02 the [Mn/Fe] uncertainties were increased by the discrepancy between
 published solar photospheric and meteoritic values.  While we believe that the techniques
 employed in this work are superior to SM02 and McWilliam et al. (2003)
 it is still possible that the [Mn/Fe] discrepancy could be due to the relatively low S/N
 of the current spectra.  Thus, it is necessary to revisit the [Mn/Fe] trend in Sgr 
 with a more extensive and higher S/N chemical abundance study.  This is particularly relevant
 following the discovery of Mn deficiencies in Omega Cen by Cunha et al. (2010).

 Although our neutron-capture element abundances show the same character as previous studies
 by Smecker-Hane \& McWilliam (2002) and Sbordone et al. (2007), there are differences
 between the results.
 In this work the [La/Y] ratios are significantly higher than those of SM02, especially at
[Fe/H]=$-$0.09 dex for which our value is higher by 0.5 dex; however, our results are slightly
lower than the [La/Y] values found by B00 and S07 (see Figures~\ref{fig-layfeh}
and \ref{fig-laylah}).  Because of limited spectral coverage, SM02 were forced
to use less-than-optimal Y~II lines, including the line at 5402.8\AA\ , which is blended
and required spectrum synthesis to account for the contamination.  For this reason, we did
not employ this Y~II line in the current work; it is possible that the blended Y~II lines
in SM02 might explain the discrepancy with all other studies.  
A comparison of published [La/Y] ratios for Sgr, in Figure~\ref{fig-layfeh}, indicates a steep
increase of [La/Y] with increasing [Fe/H].  Whether [La/Y] is flat with [Fe/H], as found by SM02, 
or steeply rising, has implications for the chemical evolution of Sgr; therefore, this [hs/ls] ratio
requires further investigation.

Notwithstanding, the prevailing [La/Y] versus [La/H] trend in Figure~\ref{fig-laylah}
shows that the locus of the maximum theoretically predicted [La/Y] AGB yields, from Cristallo et al. (2011), 
falls $\sim$0.3 dex below the measured values for Sgr from this work, B00 and S07.  This may indicate a
factor of two underestimate for the Standard Pocket, ST, of hydrogen ingested into the inter-shell region 
of the AGB star, in the theory employed by Cristallo and collaborators (going back to Gallino et al. 1998).
Curiously, the SM02 [La/Y] ratios are not in conflict with theoretical AGB s-process nucleosynthesis
predictions.  Another difficulty is that the Cristallo et al. (2011) [hs/ls] predictions showed
good agreement with chemical abundance measurements of s-process enhanced stars in the MW.  Thus, if the 
high [La/Y] ratios in Sgr prevail, then it may appear that AGB s-process is different in the MW and Sgr.
It seems possible that such differences could due to the distinct chemical composition of these two
galaxies (e.g., O, Mg, and s-process).  However, an alternate possibility is that the MW comparisons used by
Cristallo et al. (2011) did not measure the final AGB [hs/ls] ratios, either because the MW stars were 
not-yet-dead AGB stars or involved mass-transfer from AGB stars before the terminal s-process ratios were
achieved.


Small discrepancies exist for [La/Eu] in Sgr, which are slightly higher here than
in SM02 (see Figure~\ref{fig-laeulah}).  However, the differences are well within
the measurement uncertainties; we suspect that this is due to slightly high Eu
abundances in SM02, possibly as a result of the gravities employed.  We prefer
the [La/Eu] results presented here over SM02.

%
%
%

We find that the [O/Fe] trend with [Fe/H] is deficient by 0.4 dex in Sgr, relative to the MW disk, while
the trend of [Eu/Fe] and particularly the pure r-process [Eu/Fe]$_r$ (i.e., corrected for s-process Eu),
shows no deficiency compared to the MW disk.  Thus, it is not possible to explain the premature decline
in [O/Fe] with [Fe/H] by the addition of extra iron in the SNIa time-delay scenario of Tinsley (1979)
without then predicting large deficiencies in the trend of [Eu/Fe]$_r$ with [Fe/H].
A more reasonable reasonable explanation is that the deficient [O/Fe] ratios resulted from a
paucity of the highest mass SNII in Sgr; this modification of the IMF of SNII progenitors is qualitatively
consistent with the measured ratios of hydrostatic to explosive elements in Sgr.  

 We also found 0.3--0.4 dex
 enhancements in the r-process corrected [Eu/O]$_r$ and [Eu/Mg]$_r$ ratios in Sgr relative to the MW disks.
 This result simultaneously indicates that the most massive SNII progenitors were deficient in Sgr
 (i.e., either a top-light IMF or a steep IMF) AND that the r-process SNII are of lower mass than
 the oxygen-producing SNII.  Since oxygen is produced in SNII of $\sim$30 M$_{\odot}$ and above 
(e.g., Woosley \& Weaver 1995), r-process SNII are less massive than $\sim$30 M$_{\odot}$, in qualitative
 agreement with some theoretical predictions, e.g., Wanjo et al. (2003).  Deficiencies in the abundances
 of other elements made by SNII (e.g., Al, Na, the alpha-elements, and Cu) are qualitatively consistent
 with a top-light or steep IMF; however, the interpretation is complicated by the unknown nucleosynthetic
 contribution from SNIa, which depends on age and SFR.

 Similar enhancements of [Eu/O] and/or [Eu/Mg] ratios, and differences between [O/Fe] or [Mg/Fe] and 
[Eu/Fe] trends, have been seen in other dwarf galaxies, such as the LMC, Fornax and
 IC~1613; this indicates that the same deficit of high-mass SNII, suggesting a top-light or steep IMF, has
 occurred in those systems.  Such steep galaxy IGIMF slopes in dwarf galaxies have been predicted 
(e.g., Weidner \& Kroupa 2005).
The metal-poor dwarf galaxies studied to date, e.g., Sculptor, Carina, do not show enhanced
[Eu/O] and [Eu/Mg] ratios.  This might be consistent with the predicted metallicity-dependence
of the IGIMF by  Kroupa et al. (2011).  Our r-process [Eu/O]$_r$ and [Eu/Mg]$_r$ measurements
provide an important observational constraint on both the site of the r-process and the IMF 
of Sgr.

We report the first abundance measurement of Rb and the [Rb/Zr] ratio in
Sgr.  While all three of our stars have very low [Rb/Zr] ratios, our most metal-rich 
star, \#247 at [Fe/H]=$-$0.09 dex, has a remarkably low value, at [Rb/Zr]=$-$0.72 dex;
this is probably the lowest value ever reported.  Theoretical 
predictions show that this ratio could have been produced by AGB stars with [Fe/H]$\simgt$$-$1,
but not from AGB stars with [Fe/H] characteristic of M54 or a metal-poor halo.  

Comparison with the theoretical predictions indicate that the low [Rb/Zr] ratios measured in 
this work were probably produced by low-mass AGB stars ($\simlt$2 M$_{\odot}$), via the
$^{13}$C($\alpha$,n)$^{16}$O neutron source, in the main Sgr population with [Fe/H] peak near $-$0.6 dex.
Thus, the role of intermediate-mass AGB stars, relative
to low-mass AGB stars, was much diminished in Sgr compared to the MW disk and halo.  

Although published yields are insufficient to place tight constraints on the AGB masses 
indicated by the [Rb/Zr] ratios, our measurements are consistent with published yields 
for AGB stars in the range 1.3--2.0 M$_{\odot}$.  This suggests a timescale consistent with 
the ages of the 4--6 Gyr old and the 2.3 Gyr old sub-populations in Sgr, identified by
Siegel et al. (2007).

It is interesting that the [Rb/Zr] ratios in our sample of 3 stars are linearly dependent on
the [Fe/H]; while this may simply result from low-number statistics, it suggests that either 
there was a gradual decrease in [Rb/Zr], due to a steady increase in AGB metallicities
or a decrease in the mean mass of the AGB stars.  It is also possible that the
steady decline of [Rb/Zr] with [Fe/H] is due to dilution of metal-poor, high-[Rb/Zr]
material, with metal-rich, low-[Rb/Zr], ejecta from AGB stars.  More data are
required to investigate these possibilities.

The only known stellar population with similar low [Rb/Zr] values is $\omega$~Cen 
(Smith et al. 2000).  It is clear that, once again, Sgr and $\omega$~Cen are chemically 
similar, suggesting similar chemical evolution histories.  It is also clear that these 
systems provide useful grounds for testing theoretical models of the s-process in low-mass 
AGB stars.


\acknowledgements
\centerline{\bf\large Acknowledgements}
Andrew McWilliam would like to thank George Preston, Sara Bisterzo and Sergio Cristallo for useful 
conversations.  We also express deep gratitude to Carnegie Observatories librarian, John Grula, for
his efforts in obtaining copies of papers with atomic data critical to this work.


\newpage

\appendix
\section{\\I. Error Analysis}

In this work we have computed the uncertainties for our abundances following the method outlined
in the appendix of McWilliam et al. (1995); however, we have extended the formulae to include
uncertainties arising from the model atmosphere metallicity.  We present equations A1--A5
employed here.

\begin{eqnarray}\label{A1}
 \sigma (\bar{\varepsilon})^2  = \sigma_r(\bar{\varepsilon} )^2 +
                      \left(\frac{\partial \bar{\varepsilon}}{\partial T}\right)^2   \sigma_{T}^2  +
                      \left(\frac{\partial \bar{\varepsilon}}{\partial g}\right)^2   \sigma_{g}^2 +
                      \left(\frac{\partial \bar{\varepsilon}}{\partial \xi}\right)^2    \sigma_{\xi}^2 +  
                      \left(\frac{\partial \bar{\varepsilon}}{\partial [M/H]}\right)^2  \sigma_{[M/H]}^2 +  \nonumber \\
             2 \left[ \left(\frac{\partial \bar{\varepsilon}}{\partial T}\right)  
                      \left(\frac{\partial \bar{\varepsilon}}{\partial g}\right) \sigma_{Tg}  +  
                           \left(\frac{\partial \bar{\varepsilon}}{\partial T}\right)                     
                           \left(\frac{\partial \bar{\varepsilon}}{\partial \xi}\right)  \sigma_{T\xi} +   
                           \left(\frac{\partial \bar{\varepsilon}}{\partial g}\right)                     
                           \left(\frac{\partial \bar{\varepsilon}}{\partial \xi}\right)  \sigma_{g\xi} +  
                           \left(\frac{\partial \bar{\varepsilon}}{\partial [M/H]}\right)                
                           \left(\frac{\partial \bar{\varepsilon}}{\partial T}\right) \sigma_{T[M/H]} \right]
\end{eqnarray}

Where $\sigma_r(\bar{\varepsilon})$ is the uncertainty on the average abundance due to random errors,
$\sigma_X^2$ is the variance of atmosphere parameter X and $\sigma_{XY}$ is the covariance between
two atmosphere prameters, X and Y.  For the random component, we have employed the measured error on
the average abundance for species with multiple lines to estimate $\sigma_r(\bar{\varepsilon})$;
this will include the effects of unaccounted blends.  For species represented by only one line
we compted $\sigma_r(\bar{\varepsilon})$ from the uncertainty in the measured EW, according to
Equations A2 and A3, below.  
In principle, Equations A2 and A3 could be employed to compute $\sigma_r(\bar{\varepsilon})$
for all species, starting from the S/N of the observed spectrum.  Note that Equation A3 could be extended
to include other sources of uncertainty on individual line abundances, for example due to error in the
adopted damping constants.

\begin{eqnarray}\label{A1b}
\frac{1}{\sigma_r(\bar{\varepsilon} )^2} = \sum\limits^{N_{lines}} \frac{1}{\sigma_i(\varepsilon )^2}
\end{eqnarray}

\begin{eqnarray}\label{A1c}
\sigma_i (\varepsilon )^2 = 
\left(\frac{\partial \varepsilon_i}{\partial EW}\right)^2   \sigma_{EW_i}^2  +
\sigma_{log\,gf_i}^2
\end{eqnarray}

The uncertainty on the abundance ratio [A/B] is given by:

\begin{equation}\label{A2}
   \sigma(\left[A/B\right])^2 = \sigma_{\varepsilon(A)}^2 + \sigma_{\varepsilon(B)}^2 - 2\sigma_{A,B}
\end{equation}

Where the covariance between abundances of species A and B, $\sigma_{A,B}$ is given by:

\begin{eqnarray}\label{A3}
           \sigma_{A,B} =  \left(\frac{\partial \bar{\varepsilon}_A}{\partial T}\right) \left(\frac{\partial \bar{\varepsilon}_B}{\partial T}\right) \sigma_T^2 +
                 \left(\frac{\partial \bar{\varepsilon}_A}{\partial g}\right)     \left(\frac{\partial \bar{\varepsilon}_B}{\partial g}\right) \sigma_g^2 +          
                 \left(\frac{\partial \bar{\varepsilon}_A}{\partial \xi}\right)   \left(\frac{\partial \bar{\varepsilon}_B}{\partial \xi}\right) \sigma_{\xi}^2 + 
                 \left(\frac{\partial \bar{\varepsilon}_A}{\partial [M/H]}\right) \left(\frac{\partial \bar{\varepsilon}_B}{\partial [M/H]}\right)  \sigma_{[M/H]}^2 +   \nonumber \\
        \left[   \left(\frac{\partial \bar{\varepsilon}_A}{\partial T}\right)     \left(\frac{\partial \bar{\varepsilon}_B}{\partial \xi}\right) + 
                 \left(\frac{\partial \bar{\varepsilon}_A}{\partial \xi}\right)   \left(\frac{\partial \bar{\varepsilon}_B}{\partial T}\right)       \right]   \sigma_{T\xi} + 
        \left[   \left(\frac{\partial \bar{\varepsilon}_A}{\partial T}\right)     \left(\frac{\partial \bar{\varepsilon}_B}{\partial g}\right)  + 
                 \left(\frac{\partial \bar{\varepsilon}_A}{\partial g}\right)     \left(\frac{\partial \bar{\varepsilon}_B}{\partial T}\right)       \right]   \sigma_{Tg} +  \nonumber \\  
        \left[   \left(\frac{\partial \bar{\varepsilon}_A}{\partial \xi}\right)   \left(\frac{\partial \bar{\varepsilon}_B}{\partial g}\right) + 
                 \left(\frac{\partial \bar{\varepsilon}_A}{\partial g}\right)     \left(\frac{\partial \bar{\varepsilon}_B}{\partial \xi}\right)     \right]  \sigma_{g\xi} + 
        \left[   \left(\frac{\partial \bar{\varepsilon}_A}{\partial T}\right)     \left(\frac{\partial \bar{\varepsilon}_B}{\partial [M/H]}\right) + 
                 \left(\frac{\partial \bar{\varepsilon}_A}{\partial [M/H]}\right) \left(\frac{\partial \bar{\varepsilon}_B}{\partial T}\right)       \right]  \sigma_{T[M/H]}
\end{eqnarray}

In these equations we include the covariance of [M/H] with T$_{\rm eff}$, since [M/H] was based on the
Fe~I abundance which is temperature-dependent; however, we have omitted all other parameter covariances
with [M/H].  Also, we ignore the possibility of covariance between EWs of species A and B, which
holds for separated lines; however, if lines of A and B are blended together an EW covariance could exist.

The sensitivity of the element abundances to model atmosphere parameters and the model
atmosphere variances and covariances are necessary ingredients to equations A1--A5.  From
numerical experiments we have computed the dependence of abundance on atmosphere parameters
for all elements measured in star 242; the results are listed in Table~\ref{tab-dabunddpar}.

In the evaluation of our abundance measurement uncertainties we first consider
the effective temperature, T$_{\rm eff}$.
The root mean square differences of the 9 color T$_{\rm eff}$ values in Table~\ref{tab-photteff} 
is 36K, indicating an uncertainty iof 26K for each average color-temperature.  To this it was necessary
to add, in quadrature, a 15K uncertainty resulting from 0.03 magnitudes uncertainty for the V-band photometry,
common to all our colors, a 20K uncertainty due to 0.02 magnitude reddening uncertainty for Sgr
(Layden \& Sarajedini 2000), the uncertainty on the Arcturus physical T$_{\rm eff}$ value of $\pm$29K
(Koch \& McWilliam 2008), and the uncertainty of $\pm$7K on the effective temperature of the sun.
These terms indicate a total 1$\sigma$ T$_{\rm eff}$ uncertainty on our Sgr temperatures (relative to the 
solar temperature) of 47K.

\begin{deluxetable}{lcccccccc}[b]
\tabletypesize{\scriptsize}
\tablecaption{Abundance Sensitivity to Atmosphere Parameters}
\tablewidth{0pt}
\tablehead{
\colhead{} &
\multicolumn{2}{c}{$\Delta$ T$_{\rm eff}$} &
\multicolumn{2}{c}{$\Delta$ log\,g} &
\multicolumn{2}{c}{$\Delta$ $\xi$} &
\multicolumn{2}{c}{$\Delta$ [M/H]} \\
\colhead{} &
\colhead{$+$50K} & \colhead{$-$50K} &
\colhead{$+$0.2 dex} & \colhead{$-$0.2 dex} &
\colhead{$+$0.3 km\/s} & \colhead{$-$0.3 km\/s} &
\colhead{$+$0.1 dex} & \colhead{$-$0.1 dex} \\
}
\startdata
Fe~I   &  $-$0.03 &  $+$0.02     &        $+$0.04 &  $-$0.04  &           $-$0.11 &  $+$0.13   &          $+$0.03 & $-$0.01  \\
Fe~II  &  $-$0.11 &  $+$0.11     &        $+$0.17 &  $-$0.08  &           $-$0.07 &  $+$0.08   &          $+$0.08 & $-$0.01  \\
{[O~I]a} &  $+$0.01 &  $+$0.00     &        $+$0.07 &  $-$0.05  &           $-$0.01 &  $+$0.02   &          $+$0.03 & $-$0.02  \\
Na~I   &  $+$0.05 &  $-$0.04     &        $+$0.01 &  $-$0.01  &           $-$0.05 &  $+$0.07   &          $+$0.01 & $+$0.01  \\
Mg~I   &  $-$0.02 &  $+$0.03     &        $+$0.05 &  $-$0.01  &           $-$0.06 &  $+$0.07   &          $+$0.03 & $+$0.00  \\
Al~I   &  $+$0.03 &  $-$0.03     &        $+$0.01 &  $+$0.00  &           $-$0.04 &  $+$0.05   &          $+$0.01 & $+$0.01  \\
Si~I   &  $-$0.06 &  $+$0.07     &        $+$0.09 &  $-$0.03  &           $-$0.04 &  $+$0.04   &          $+$0.05 & $+$0.00  \\
Ca~I   &  $+$0.05 &  $-$0.05     &        $+$0.01 &  $+$0.00  &           $-$0.14 &  $+$0.18   &          $+$0.01 & $+$0.00  \\
Ti~I   &  $+$0.07 &  $-$0.06     &        $+$0.02 &  $-$0.02  &           $-$0.20 &  $+$0.26   &          $+$0.02 & $-$0.01  \\
Ti~II  &  $-$0.04 &  $+$0.03     &        $+$0.11 &  $-$0.08  &           $-$0.05 &  $+$0.05   &          $+$0.05 & $-$0.03  \\
V~I    &  $+$0.07 &  $-$0.06     &        $+$0.03 &  $-$0.03  &           $-$0.18 &  $+$0.19   &          $+$0.02 & $-$0.02  \\
Mn~I   &  $+$0.01 &  $-$0.00     &        $+$0.06 &  $-$0.03  &           $-$0.10 &  $+$0.12   &          $+$0.03 & $-$0.01  \\
Cu~I   &  $+$0.00 &  $+$0.00     &        $+$0.08 &  $-$0.05  &           $-$0.08 &  $+$0.10   &          $+$0.05 & $-$0.01  \\
Rb~I   &  $+$0.07 &  $-$0.07     &        $+$0.00 &  $+$0.00  &           $-$0.01 &  $+$0.01   &          $+$0.01 & $+$0.00  \\
Zr~I   &  $+$0.08 &  $-$0.08     &        $+$0.03 &  $-$0.03  &           $-$0.18 &  $+$0.27   &          $+$0.02 & $-$0.02  \\
Y~II   &  $-$0.01 &  $+$0.01     &        $+$0.09 &  $-$0.07  &           $-$0.03 &  $+$0.05   &          $+$0.05 & $-$0.03  \\
La~II  &  $+$0.01 &  $-$0.01     &        $+$0.09 &  $-$0.08  &           $-$0.05 &  $+$0.05   &          $+$0.05 & $-$0.03  \\
Eu~II  &  $-$0.01 &  $+$0.01     &        $+$0.10 &  $-$0.07  &           $-$0.03 &  $+$0.04   &          $+$0.05 & $-$0.02  \\
\enddata
\tablecomments{a-- oxygen abundances assume RGB carbon depletion of 0.2 dex.  If $\Delta$C= +0.10 dex  $\Delta$O = +0.04 dex
If $\Delta$C= -0.10 dex  $\Delta$O = -0.03 dex.}
\label{tab-dabunddpar}
\end{deluxetable}

In this work we have adopted log\,g values from the BASTI Ischrones, using the adopted T$_{\rm eff}$
and metallicity, [M/H], in an iterative differential abundance analysis.  Our formal
uncertainty in log\,g results, directly, from the temperature uncertainty.  We computed the
log\,g uncertainty from the slope of log\,g versus T$_{\rm eff}$ times $\sigma$T; $\sigma$(T)=47K
corresponds to $\sigma$(log\,g)=0.040 dex.

Because T$_{\rm eff}$ and our adopted log\,g values are correlated it is necessary to include the
covariance between these two parameters, $\sigma_{Tg}$, when evaluating the abundance uncertainties. 
The covariance between temperature and gravity, $\sigma_{\rm Tg}$ was computed using Equation~A6:

\begin{equation}
  \sigma_{\rm Tg} = <\Delta g \Delta T> =  \left(\frac{\partial g}{\partial T}\right) \sigma^2_T
\end{equation}

Where the gradient, $\left({\partial g}/{\partial T}\right)$, was measured from the BASTI isochrone
appropriate for Star~242.  Accordingly, we find $\sigma_{\rm Tg}$=1.88.
We remind the reader that covariances between atmosphere parameters depend upon the method
used to measure them.  We have employed photometric temperatures and gravities with help
from theoretical isochrones; $\sigma_{\rm Tg}$ would have been different for spectroscopically
determined gravities and/or temperatures.

The standard error in the slope of the plot of [Fe~I/H] versus EW for Star~242
corresponds to a 1$\sigma$ microturbulent velocity uncertainty of 0.02 km/s.  Because 
we have employed a line by line differential analysis, the scatter in this slope is
particularly small, since there is no dispersion due to uncertainties in $gf$ values.
For the covariance between gravity and microturbulent velocity, $\sigma$$_{g\xi}$,
we estimated the maximum change in slope in the $\epsilon$(Fe~I) versus EW plot
for a change in gravity of 0.2 dex, thus permitting calculation of $<\Delta g\Delta\xi>$.
Accordingly, we found $\sigma$$_{T\xi}$=$-$0.0011; this covariance estimate would have
been larger had absolute $gf$ values been used in the analysis.  For $\sigma_{[M/H]}$
we adopted the uncertainty in [Fe~I/H], which is dominated by the uncertainty
in T$_{\rm eff}$.  Clearly, the exact chemical composition, amongst other variables, could affect
our choice of model atmosphere metallicity; however, we note that increasing $\sigma_{[M/H]}$ to 0.06
dex made a negligible difference to the total error budget.  For the temperature-metallicity 
covariance, $\sigma_{T[M/H]}$, we employed the abundance sensitivities in Table~\ref{tab-dabunddpar},
resulting in a covariance of 1.105 K dex.  Finally, we could see no significant covariance between
temperature and microturbulent velocity, so this has been assumed to be zero in this work.
Our resultant standard errors and covariances are summarized in Table~\ref{tab-varcovar} below.

\begin{deluxetable}{ll}
\tabletypesize{\scriptsize}
\tablecaption{Atmosphere Parameter Variances and Covariances}
\tablewidth{120pt}
\startdata
\tableline \\
$\sigma_{T}$        &    47.0       \\
$\sigma_{g}$        &    0.040      \\
$\sigma_{\xi}$      &    0.02       \\
$\sigma_{[M/H]}$    &    0.024      \\
                    &               \\
$\sigma_{Tg}$       &    1.88       \\
$\sigma_{T[M/H]}$   &    1.105      \\
$\sigma_{g\xi}$     & \llap{$-$}0.0011     \\
$\sigma_{T\xi}$     &    0.00       \\
\enddata
\label{tab-varcovar}
\end{deluxetable}

When the variances and covariances from Table~\ref{tab-varcovar} are inserted
into equations A1--A3, and combined with the random errors of
Table~\ref{tab-abunds} to estimate the EW measurement uncertainties,
we compute the abundance uncertainties, relative to H, Fe~I and Fe~II, as shown
in Table~\ref{tab-finalerrors}.  It is notable, and unexpected, that the ratios [X/Fe~II] are
always inferior to [X/Fe~I] in Table~\ref{tab-finalerrors}, even for [O I],
Ti~II and other ionized lines; we assume that this is because our stellar 
temperatures are cool enough that Fe~I is the dominant ionization stage, unlike 
the red giant stars a few hundred degrees hotter.

\begin{deluxetable}{lcccccc}[h]
\tabletypesize{\scriptsize}
\tablecaption{Abundance Ratio Uncertainties}
\tablewidth{0pt}
\tablehead{\\
&
\multicolumn{3}{c}{\underline{Atmosphere Uncertainties}} &
\colhead{} &
\colhead{} \\ [5pt]
\colhead{Ion} & 
\colhead{$\sigma$[X/H]} &
\colhead{$\sigma$[X/Fe\,I]} &
\colhead{$\sigma$[X/Fe\,II]} &
\colhead{$\sigma_{rand}[X/H]^{\rm a}$}&  
\colhead{$\sigma_{\rm total}$[X/Fe\,I]}&
\colhead{$\sigma_{\rm total}$[X/Fe\,II]}   \\
}
\startdata
Fe~I   &     0.02    &        ...    &     0.06   &             0.023           &        0.03\rlap{$^{\rm b}$} &    ...       \\
Fe~II  &     0.07    &        0.06   &     ...    &             0.105           &        ...     &    0.13\rlap{$^{\rm b}$}  \\
{[O~I]} &    0.02    &        0.03   &     0.09   &             0.033\rlap{$^{\rm c}$}  &        0.04     &    0.10      \\
Na~I   &     0.05    &        0.06   &     0.11   &             0.13\phantom{0} &        0.14             &    0.17      \\
Mg~I   &     0.02    &        0.01   &     0.05   &             0.015           &        0.02             &    0.05      \\
Al~I   &     0.03    &        0.04   &     0.10   &             0.035           &        0.05             &    0.11      \\
Si~I   &     0.05    &        0.03   &     0.03   &             0.065           &        0.07             &    0.07      \\
Ca~I   &     0.05    &        0.06   &     0.12   &             0.072           &        0.09             &    0.14      \\
Ti~I   &     0.07    &        0.08   &     0.15   &             0.064           &        0.10             &    0.16      \\
Ti~II  &     0.02    &        0.00   &     0.06   &             0.20\phantom{0} &        0.20             &    0.21      \\
V~I    &     0.08    &        0.08   &     0.14   &             0.057           &        0.10             &    0.15      \\
Mn~I   &     0.03    &        0.03   &     0.09   &             0.046           &        0.05             &    0.10      \\
Cu~I   &     0.03    &        0.03   &     0.09   &             0.096\rlap{$^{\rm c}$}  &        0.10     &    0.13      \\
Rb~I   &     0.07    &        0.08   &     0.14   &             0.01\phantom{0} &        0.08             &    0.14      \\
Zr~I   &     0.09    &        0.10   &     0.15   &             0.045           &        0.11             &    0.16      \\
Y~II   &     0.02    &        0.03   &     0.08   &             0.050           &        0.06             &    0.09      \\
La~II  &     0.04    &        0.05   &     0.10   &             0.092           &        0.10             &    0.14      \\
Eu~II  &     0.02    &        0.03   &     0.08   &             0.058\rlap{$^{\rm c}$}  &        0.07     &    0.10      \\
\enddata
\tablecomments{a: For species with more than one measured line, random abundance errors, due to EW
                  uncertainties, were adopted from the error on the mean abundances, derived from 
                  the dispersions in Table~\ref{tab-abunds}.\quad\quad
               b: Instead of $\sigma$$_{rm total}$[X/Fe] we provide $\sigma$$_{rm total}$[Fe/H] ratios.\quad\quad
               c: Random abundance errors for O, Cu and Eu were computed from 1$\sigma$ EW measurement
                  uncertainties, based on the S/N for each pixel of each line, at 4.6, 4.2 and 7.0 m\AA\
                  respectively.}
\label{tab-finalerrors}
\end{deluxetable}

Caveats regarding our treatment of the abundance measurement uncertainty are related to
systematic effects, particularly the omission of physics in the one-dimensional LTE model
atmospheres and the use of LTE radiative transfer.  Improvements could include
3-dimensional hydrodynamical atmospheres with non-LTE radiative
transfer.  However, it is hoped that our differential abundance technique has 
minimized the deviations from these omissions; also, abundance corrections
from these two effects tend to have opposite sign, cancelling to some degree.
We also warn that we have assumed that our stars are on the RGB, rather than AGB,
when selecting the gravity for our model atmospheres.  On the RGB our stars have log\,g
values only 0.07 dex higher than the AGB.  However, if the mass-loss prescription
used in the BASTI isochrones is greatly different this gravity difference may be
somewhat larger or smaller.  Another issue is that we have employed scaled-solar composition
model atmospheres, from the Kurucz grid, whereas, our analysis shows that elements that 
are important electron donors, namely Na, Mg, Al, and Si are deficient in all our stars.
A simple calculation for locations in one of our model atmospheres indicates
that N$_e$ is reduced by $\sim$30\%, or $\sim$0.1 dex.  Such changes in N$_e$
should reduce the abundances of Fe~II, Ti~II and [O~I] lines.  A more
complete propagation of errors would consider this effect.  For this reason, although
or results suggest a reduction in the measurement error for [O~I] and Ti~II by
ratioing with Fe~I abundances, we prefer to ratio with Fe~II abundances, in order
to subtract out systematic abundance errors due to inappropriate N$_e$ in the model 
atmospheres.

\vfill\eject

\section{II.  HFS References}



\begin{deluxetable}{ll}[b]
\centering
\tablecaption{HFS References}
\tablewidth{0pt}
\tablehead{\\
\colhead{Ion} &
\colhead{References} \\
}
\startdata
V~I    &   Kopfermann, H. \& Rasmussen, E. 1936, Z. Phys., 98, 624  \\
       &  \\
       &   Childs,W.J., Poulsen, O., Goodman, L.S., \& Crosswhite, H. 1979, Phys. Rev.A19, 168  \\
       &  \\
       &   Kurucz, R.L., unpublished (http://kurucz.harvard.edu/)  \\
       &  \\
       &   Johnson, J.A., Ivans, I.I., and Stetson, P.B. 2006, \apj, 640, 801   \\
       &  \\
Mn~I   &   Handrich, E., Steudel, A., \& Walther, H. 1969, Phys. Lett., 29A, 486  \\
       &   (components computed by R.L. Kurucz, http://kurucz.harvard.edu/)  \\
       &  \\
Cu~I   &   Cunha, K., Smith, V., Suntzeff, N.B., Norris, J.E., Ca Costa, G, \& Plez, B. 2002,  \\
       &   \aj, 124, 379 (see Biehl 1976)  \\
       &  \\
       &   Biehl, D. 1976, Ph.D. thesis, Univ. Kiel  \\
       &  \\
Rb~I   &   Daniel A. Steck, ``Rubidium 85 D line Data'', available online at  \\
       &   http://steck.us/alkalidata (revision 2.1.4, 23 December 2010)  \\
       &  \\
       &   Daniel A. Steck, ``Rubidium 87 D line Data'', available online at  \\
       &   http://steck.us/alkalidata (revision 2.1.4, 23 December 2010)  \\
       &  \\
       &   Banerjee, A., Das, A., \& Natarajan, V. 2004, Europhys. Lett., 65 (2), pp.172-178  \\
       &  \\
Y~II   &   Persson, J.R. 1997, Z. Phys. D., 42, 259 \\
       &  \\
       &   Beck, D.R. 1992, Phys. Rev. A, 65, 1399 \\
       &  \\
       &   Dinneen, T.P., Mansour, N.B., Kurtz, C., \& Young, L.  1991, Phys. Rev. A, 43, 4824 \\
       &  \\
       &   Villemoes, P., Arnesen, A., Heijkenskj\"old, F., Kastberg, A., \& Larsson, .O. 1992, \\
       &   Physica Scripta, 46, 45 \\
       &  \\
Zr~I   &   Chevalier, G., Gagn\'e, J.-M., \& Pianarosa, P. 1988, J. Opt. Soc. Am. B, 5, 1492 \\
       &  \\
       &   Chevalier, G., \& Gagn\'e, J.-M. 1986, Optics Communications, 57, 327 \\
       &  \\
       &   Bouazza, S., Gough, D.S., Hannaford, P., \& Wilson, M. 2002, J. Phys. B, 35, 2397 \\
       &  \\
       &   Gough, D.S., \& Hannaford, P. 1988, Optics Communications, 67, 209 \\
       &  \\
       &   McLean, R.J., Hannaford, P., \& Larkins, P.L. 1993, Optics Commumications, 102, 43 \\
       &  \\
       &   B\"uttgenbach, S., Dicke, R., Gebauer, H., Kuhnen, R.; Tr\"aber, F. 1978,  Z. Phys. A, 286, 125 \\
       &  \\
La~II  &   Lawler, J.E., Bonvallet, G., \& Sneden, C. 2001, \apj, 556, 452 \\
       &  \\
       &   Furmann, B., Elantkowska, M., Stefa\'nska, D., Ruczkowski, J., \& Dembczy\'nski, J.  \\
       &   2008, J. Phys. B: At. Mol. Opt. Phys., 41, 235002 \\
       &  \\
       &   Furmann, B., Ruczkowski, J., Stefa\'nska, D., Elantkowska, M., \& Dembczy\'nski, J.  \\
       &   2008, J. Phys. B: At. Mol. Opt. Phys., 41, 215004 \\
       &  \\
Eu~II  &   Lawler, J.E., Wickliffe, M.E., Den Hartog, E.A., \& Sneden, C. 2001, \apj, 563, 1075 \\
\\
\enddata
\label{tab-hfsrefs}
\end{deluxetable}
\vfill \eject

\end{document}